\documentclass[acmsmall]{acmart}
\AtBeginDocument{%
  \providecommand\BibTeX{{%
    \normalfont B\kern-0.5em{\scshape i\kern-0.25em b}\kern-0.8em\TeX}}}

\setcopyright{acmcopyright}
\copyrightyear{2024}
\acmYear{2024}
\acmDOI{10.1145/1122445.1122456}

\acmJournal{TOSEM}
\acmVolume{37}
\acmNumber{4}
\acmArticle{111}
\acmMonth{1}

\usepackage{microtype}
\usepackage{textcomp}
\usepackage{bm}
\usepackage[caption=false]{subfig}
\usepackage{flushend}
\usepackage{csquotes}
\usepackage{booktabs} 
\usepackage{framed}
\usepackage{graphicx}
\usepackage{tabularx}
\usepackage{longtable}
\usepackage{graphicx}
\usepackage{url}
\usepackage{multirow}
\usepackage{enumitem}
\usepackage{hyperref}
\usepackage{lscape}
\usepackage[normalem]{ulem}
\useunder{\uline}{\ul}{}
\usepackage{latexsym}
\usepackage{xcolor}
\usepackage{color, colortbl}
\usepackage{hyperref}

\definecolor{g}{gray}{0.925}
\usepackage{graphics}

\definecolor{lightgray}{gray}{0.9}
 
\usepackage{siunitx}  
\usepackage{textgreek}  
\usepackage{wasysym}
\usepackage{makecell}

\usepackage{tabularx}

\usepackage{amsthm}

\newtheorem*{hypothesis}{Hypothesis}

\begin{document}

\title{Navigating the Complexity of Generative AI Adoption in Software Engineering}

\author{Daniel Russo}
\email{daniel.russo@cs.aau.dk}
\orcid{0000-0001-7253-101X}
\affiliation{%
  \institution{Department of Computer Science, Aalborg University}
  \streetaddress{A.C. Meyers Vaenge, 15, 2450}
  \city{Copenhagen}
  \country{Denmark}}

\renewcommand{\shortauthors}{Russo, 2024}

\begin{abstract}
This paper explores the adoption of Generative Artificial Intelligence (AI) tools within the domain of software engineering, focusing on the influencing factors at the individual, technological, and social levels. We applied a convergent mixed-methods approach to offer a comprehensive understanding of AI adoption dynamics. We initially conducted a questionnaire survey with 100 software engineers, drawing upon the Technology Acceptance Model (TAM), the Diffusion of Innovation Theory (DOI), and the Social Cognitive Theory (SCT) as guiding theoretical frameworks. Employing the Gioia Methodology, we derived a theoretical model of AI adoption in software engineering: the Human-AI Collaboration and Adaptation Framework (HACAF). This model was then validated using Partial Least Squares – Structural Equation Modeling (PLS-SEM) based on data from 183 software engineers. 
Findings indicate that at this early stage of AI integration, the compatibility of AI tools within existing development workflows predominantly drives their adoption, challenging conventional technology acceptance theories. The impact of perceived usefulness, social factors, and personal innovativeness seems less pronounced than expected. The study provides crucial insights for future AI tool design and offers a framework for developing effective organizational implementation strategies.
\end{abstract}



\maketitle

\section{Introduction}

The transformational promise of Artificial Intelligence (AI) is becoming increasingly evident across various sectors, with AI models demonstrating human-like competencies in areas as diverse as natural language understanding and image recognition~\cite{zhang2021ai}. One domain where this potential is particularly salient is software engineering, a critical function within contemporary organizations. This significance is underscored by the increasing pervasiveness of software in a broad range of products and services, with digital features enhancing their value~\cite{satyanarayanan2001pervasive}. 
From the initial phases of the software development lifecycle, AI tools can serve as valuable allies. Generative AI can sift through vast data sources like user feedback, market trends, and system logs, providing insights for feature ideation. During systems analysis and design, AI-enhanced tools can propose multiple IT architectural designs and swiftly adapt configurations, expediting the design process and product launches. In the coding phase, AI not only assists in generating code but also aids developers by crafting initial code drafts, swiftly detecting patterns, and serving as a knowledge repository. In the testing phase, AI tools can autonomously generate test cases and automate specific testing functions. During deployment, AI tools streamline the release process, ensuring that software versions are seamlessly integrated into existing systems, while also monitoring for potential deployment anomalies and facilitating rollback strategies if needed. When it comes to maintenance, insights derived from AI can aid software engineers in diagnosing issues, suggesting fixes, and predicting potential areas of improvement.
The implications of AI for the field of software development could be momentous, with predictions indicating a surge in productivity ranging from 20 to 45 percent~\cite{mckinsey2023EconomicPotential}. This substantial increase could be achieved by streamlining traditional tasks like crafting preliminary code drafts, refining existing code structures (refactoring), or conducting thorough root-cause analyses. The integration of AI not only reduces the time commitment for these activities but also enhances the overall efficiency and effectiveness of the software development process~\cite{peng2023impact}.
Nonetheless, despite the prospective advantages, the incorporation of language models into software engineering appears to be intricate and fraught with challenges. Indeed, there are even indications that usage of Large Language Models is on the decline, possibly as a result of end-user experimentation that found them to be ill-suited to their requirements~\cite{Economist_2023}.
Consequently, a pressing need exists to unravel the core determinants influencing the adoption of Generative AI-driven tools, such as LLM-powered tools. As we know, a diverse range of elements shape the modality and rationale behind software engineers' decision to employ language models. These incorporate both technical components, such as model quality and performance, and non-technical components, including perceived utility and ease of use~\cite{venkatesh2016unified}. 
Yet, there has been limited empirical research on the factors that influence language model adoption in software engineering. 
Hence, we formulate our research questions as follows:
\\[.1in]
\noindent\textit{\textbf{Research Question}}: \textit{
What influences the adoption of Generative AI tools in software engineering?}
\\

In our endeavor to explore our research question, we have applied a convergent mixed-methods approach, investigating the adoption of Generative AI and Large Language Models within the sphere of software engineering. 
To frame our understanding of AI adoption, we referenced three principal theoretical frameworks examining individual, technological, and social-level influences. These frameworks included the Technology Acceptance Model (TAM)~\cite{venkatesh2000theoretical}, the Diffusion of Innovation Theory (DOI)~\cite{rogers2010diffusion}, and the Social Cognitive Theory (SCT)~\cite{bandura2014social}. By incorporating these well-validated theories, we could thoroughly comprehend the determinants of language model adoption and investigate the distinct ways these variables are operationalized in the software engineering domain.
We initiated our research by conducting a questionnaire survey with a cohort of 100 software engineers. The design of these questionnaire survey was influenced by the main dimensions of our selected theories. 
The collected data was analyzed using the Gioia Methodology~\cite{gioia2013seeking}, facilitating the development of our preliminary theoretical model. 
This provisional theoretical model was then validation using Partial Least Squares -- Structural Equation Modeling (PLS-SEM), supported by data collected from 183 software engineers. The convergence of insights derived from this comprehensive and multifaceted investigation enhances our understanding of AI adoption within software engineering. Moreover, by understanding the adoption dynamics and impact of these disruptive technologies, this research holds potential to guide the design of future AI tools and offer pertinent recommendations for organization-wide implementation strategies.
Indeed, generative AI tools represent a disruptive innovation in the software engineering domain, as defined by Christensen's concept of ``disruptive innovation'' \cite{christensen1997}. These tools, while initially targeting niche applications or underserved market segments, have the potential to revolutionize traditional software engineering practices. Their ability to automate complex tasks, generate code, and offer solutions based on vast datasets challenges the status quo and can lead to a paradigm shift in how software development is approached. Over time, as these tools become more refined and widely adopted, they could displace established methodologies and tools, much like how disruptive innovations reshape industries. By understanding the adoption dynamics of such transformative technologies, this research offers insights into their potential trajectory and implications, guiding the design of future AI tools and providing recommendations for their strategic implementation across organizations.

The structure of this article unfolds as follows. In Section~\ref{sec:related}, we present a comprehensive review of related works. Our mixed-methods investigation commences with an initial theory induction, a process we detail thoroughly in Section~\ref{sec:QualInvestigation}. We then analyze the results of this process in Section~\ref{sec:results}. From these findings, we craft our theoretical framework in Section~\ref{sec:HACAF}, elucidating its hypotheses. Our model undergoes a rigorous validation process using Partial Least Squares - Structural Equation Modeling, as reported in Section~\ref{sec:QuantStudy}. In the concluding sections, we reflect on the broader implications and potential limitations of our study in Section~\ref{sec:Discussion}, and sketch out future research trajectories in Section~\ref{sec:conclusion}.

\section{Related Work}
\label{sec:related}

The software engineering landscape is undergoing a transformation with the introduction of Generative AI, especially through the capabilities of Large Language Models (LLMs). LLMs, such as the Generative Pre-trained Transformer (GPT) series by OpenAI \cite{brown2020language}, owe their prowess to the foundational role of transformer architectures in natural language processing, which have been influential for several years \cite{vaswani2017attention}. Beyond their known proficiency in generating human-like text \cite{holtzman2019curious}, LLMs, in the realm of software engineering, are poised to offer code suggestions, assist in automated documentation, aid in requirement elicitation, and more. The evolution of LLMs as Generative AI has been further propelled by the integration of transformer architectures, enhancing their understanding and generation of context \cite{kaplan2020scaling}. As we navigate the confluence of LLMs and software engineering, it becomes evident that their potential extends beyond mere text generation. They are emerging as collaborative tools, set to redefine various facets of the software development lifecycle. Hence, throughout this paper, we use the terms Generative AI and Large Language Models interchangeably, emphasizing their relevance and potential in software engineering

A multitude of academic disciplines are currently exploring the potential implications of such advanced technologies within their respective fields. This section will discuss particular context of software engineering, highlighting the prevalent themes and concerns associated with AI tools.

\subsection{Assessing Generated Code: Correctness and Quality}

A prominent strand of inquiry involves assessing the accuracy and quality of code generated by AI systems such as GitHub Copilot. Studies by Nguyen and Nadi, Dakhel et al., and Yetistiren, all conducted empirical assessments to evaluate the correctness of the code generated by Copilot, and found varying degrees of success depending on the programming language and the complexity of the task \cite{nguyen2022empirical, dakhel2023github, yetistiren2022assessing}. This shared focus on evaluation signifies the importance of assessing the functional integrity of the code generated by AI tools, which is a fundamental concern in software engineering.

\subsection{Evaluation Criteria: Diverse Approaches}

While there were commonalities in evaluating Copilot's performance, the specific aspects of evaluation varied among the studies. Nguyen and Nadi focused on the performance of Copilot across different programming languages \cite{nguyen2022empirical}, while Mastropaolo et al. investigated the robustness of Copilot in relation to semantic-preserving changes in the natural language description \cite{mastropaolo2023robustness}. Yetistiren conducted a comprehensive assessment of the generated code in terms of validity, correctness, and efficiency \cite{yetistiren2022assessing}. These differences underscore the multifaceted nature of AI code generation and the various dimensions that need assessment.

\subsection{Enhancing Code Productivity}

Productivity in software development is positively impacted by the integration of AI tools, with tools like Copilot significantly increasing the speed of code production \cite{imai2022github, peng2023impact}. While these tools are lauded for their productivity-enhancing capabilities, understanding their performance and limitations is vital to leveraging their potential effectively.
Studies by Tian et al. \cite{tian2023chatgpt} and Camara et al. \cite{camara2023assessment} shed light on the abilities and constraints of Large Language Models, such as ChatGPT. Empirical evaluation suggests that while these models demonstrate aptitude in simpler, well-structured tasks, they tend to struggle with complex tasks involving semantic nuance \cite{tian2023chatgpt}. Additionally, a significant relationship was identified between the length of the input sequence and the quality of the output, with longer inputs often leading to poorer results \cite{camara2023assessment}.
These findings highlight the nuanced role of AI in software development productivity. While they enhance speed, the complexity of tasks and the length of input sequences can serve as limiting factors, pointing to areas for further improvement and optimization in these models.

\subsection{Comparing Methods}

Distinctly, Sobania et al. embarked on a comparative study between Copilot and genetic programming, another approach in automatic program synthesis. They concluded that, despite comparable performances, genetic programming was not as mature for practical software development as Copilot \cite{sobania2022choose}. This comparative analysis provides a unique perspective on the landscape of automatic programming methodologies.

\subsection{Pedagogical Concerns}

Pedagogical Concerns have been discussed by both Wermelinger and Dakhel et al.'s studies touched upon the implications of AI tools like Copilot in educational settings. While Wermelinger explored the implications of Copilot on teaching and assessment methods in programming courses \cite{wermelinger2023using}, Dakhel et al. discussed the potential challenges for novice developers who might fail to filter Copilot's non-optimal solutions due to lack of expertise \cite{dakhel2023github}. These similarities highlight the significant pedagogical implications of integrating AI tools in education.

\subsection{AI's Influence on Software Development Process}

Concerns around the integration and security of AI tools like GitHub Copilot are a shared finding between Jaworski and Piotrkowski \cite{jaworski2023study} and Zhang et al. \cite{Zhang2023PracticesAC}. Both studies illustrate that despite the potential benefits of these tools, developers express hesitation and face challenges when incorporating them into their workflows, due to integration difficulties and security worries.
However, these studies diverge when examining developer interactions. Zhang et al. detail more practical aspects like programming languages and IDEs used with Copilot, whereas Jaworski and Piotrkowski focus on the developers' sentiment towards AI tools.
Ernst and Bavota \cite{Ernst2022AI}, although also discussing the complexities of AI integration, differ by highlighting additional challenges related to legal compliance and bias. This broadens the conversation on AI's impact on software development beyond technical aspects to include ethical and legal considerations.
Another commonality, albeit from a different angle, is found in Bird et al. \cite{bird2022taking} and Mozannar et al. \cite{mozannar2022reading}. Both studies touch on the evolving role of developers as AI tools become more pervasive. Bird et al. suggest a shift towards developers spending more time reviewing AI-generated code, whereas Mozannar et al. provide a structured analysis of developer interactions with AI tools, revealing inefficiencies and time costs.
Thus, while the studies largely converge on the transformative potential and challenges of AI tools in software development, they also bring unique perspectives to the table, expanding the discourse to include aspects like legal concerns, workflow changes, and time costs.

\subsection{Community Influence and Trust in AI Tools}

The role of the community in shaping developers' trust in AI tools is investigated by Cheng et al. \cite{cheng2022would}. They present a detailed analysis of how online communities, such as forums and discussion groups, influence developers' perception of AI tools. Their research indicates that shared experiences and collective discussions play a significant role in shaping developers' trust in AI assistance. 

\subsection{Generative AI in Non-Coding Activities}

Generative AI's impact on non-coding activities in software development is multifaceted. A prominent aspect is the surge in productivity and creativity improvements, as noted by Ozkaya \cite{ozkaya2023next}. This perspective is echoed by Schmidt \cite{schmidt2023speeding} who alludes to the potential of AI in swiftly spotting and fixing bugs. However, the two diverge in their emphasis: while Ozkaya focuses on the broader paradigm shifts in software engineering conferences and research dynamics, Schmidt stresses into the challenges of ensuring trustworthiness in AI systems, suggesting a complexity in their deployment.
This theme of trust is further expanded upon by Ebert and Louridas \cite{ebert2023generative}. They discuss the evolving nature of software systems as being more adaptive, self-modifying, and learning-oriented. Contrasting this with Ozkaya's perspective \cite{ozkaya2023next} on the implications of shifting to AI tools, Ebert and Louridas shed light on the intricate challenges in validating such systems. They suggest that traditional software testing paradigms may no longer suffice, introducing a dimension of complexity in the deployment of AI in software development.
Furthermore, the question of data quality enhancement through Generative AI offers another layer of analysis. Ebert and Louridas \cite{ebert2023generative} detail the process of fine-tuning LLMs on specific datasets, emphasizing the resultant increase in output quality. This stands in contrast with the broader shifts and challenges highlighted by Ozkaya, focusing instead on the tactical advantages offered by AI tools.
Overall, while there's a consensus on the transformative potential of Generative AI in software development, the literature also paints a picture of the challenges and nuances. From broader shifts in the research landscape to tactical advantages and validation challenges, the discourse on AI's role in non-coding activities is both rich and diverse, necessitating a holistic understanding for effective deployment.

\subsection{Usability of AI Programming Assistants}

The usability of AI programming assistants has been a focal point in the research, with the key motivations for usage identified as reduction in keystrokes, quick task completion, and syntax recall \cite{liang2023understanding,vaithilingam2022expectation}. However, developers often encounter challenges with tool control, output alignment with requirements, and difficulties with understanding, editing, and debugging generated code snippets \cite{liang2023understanding, vaithilingam2022expectation}.
For novice programmers, cognitive and metacognitive issues arise while using these tools for assignments, indicating a need for better supportive design \cite{prather2023s}. Also, developers exhibit distinct interaction modes, each requiring different forms of tool support \cite{barke2023grounded}.
This signifies a necessity for usability improvements in AI programming assistants, focusing on user control, cognitive effort minimization, and support for interaction modes.
\\

In sum, the review of related work underscores the transformative potential and multifaceted challenges posed by Large Language Models in the realm of software engineering. The corpus of research spans areas such as the evaluation of generated code's accuracy and quality, the augmentation of productivity, contrasting methodologies, pedagogical implications, AI's influence on software development processes, the role of community in fostering trust, and the usability of AI programming assistants. Each study contributes uniquely to our understanding of AI's role in software engineering, highlighting the complexity of the issues at hand. While the field has made significant strides in leveraging AI's potential, the need for robust evaluation, tailored usability, and mindful integration into educational and professional settings is a recurring theme.

Beyond the aforementioned research, it is crucial to note the existence of other code generators besides Copilot. Tools such as Alphacode \cite{li2023alphacode}, Amazon Codewhisperer \cite{aws2023codewhisperer}, BlackBox AI \cite{blackbox2023}, CodeComplete \cite{codecomplete2023}, CodeGeeX \cite{zheng2023codegeex}, Codeium \cite{codeium2023}, Mutable AI \cite{mutable2023}, GhostWriter Replit \cite{replit2023ghostwriter}, and Tabnine \cite{tabnine2023} also play roles in the domain of AI-powered code generation. Interestingly, while these tools are acknowledged in the landscape, there is a glaring lack of empirical research evaluating them. Our research of the literature revealed only three papers that even mention these tools, and solely within the context of related work or discussion sections \cite{cheng2022would,wang2023investigating,gozalo2023survey}. This presents a clear gap in the current body of research.

More specifically, while existing research thoroughly investigates the performance, usability, and impact of Generative AI tools in software engineering, it primarily focuses on the tools themselves, largely overlooking the factors influencing their adoption. A notable exception is Cheng et al.'s \cite{cheng2022would} exploration of community influence on developers' trust. Yet, this is only one facet of the broader adoption landscape, which includes individual, organizational, technological, and environmental factors. Our research question addresses this clear gap in the literature, aiming to provide a comprehensive understanding of the factors driving or hindering the adoption of these tools.

\section{Theory Generation}
\label{sec:QualInvestigation}

\subsection{Theoretical foundation}

The multifaceted nature of technology adoption demands the application of comprehensive theoretical frameworks that can sufficiently capture and explain the influencing factors. The Technology Acceptance Model, Diffusion of Innovation Theory, and Social Cognitive Theory together offer a robust approach towards understanding the complexity of language model adoption in software engineering.

\subsubsection{Technology Acceptance Model (TAM)}
TAM has been widely acclaimed for its relevance and efficiency in predicting and explaining the acceptance of various forms of technology~\cite{venkatesh2000theoretical}. Its core constructs—perceived usefulness and perceived ease of use—serve as an excellent starting point for understanding adoption behavior. For instance, if language models are perceived as beneficial and easy to use, software engineers are more likely to embrace them. Given its robustness and simplicity, TAM provides a foundation for understanding the fundamental determinants of technology adoption and aids in diagnosing the basic barriers to language model adoption.

\subsubsection{Diffusion of Innovation Theory (DOI)}
While TAM primarily focuses on user perceptions, DOI complements TAM by addressing the technological characteristics influencing adoption. Rogers~\cite{rogers2010diffusion} identified key attributes of innovations—relative advantage, compatibility, complexity, trialability, and observability—that significantly affect their adoption rates. As an innovation in software engineering, the acceptance of language models could be shaped by these attributes. For example, the relative advantage of language models over traditional programming methods could be a strong motivator for adoption. The compatibility of language models with existing practices and the complexity of these models might also play crucial roles. Thus, DOI adds depth to our understanding of technology-specific factors influencing language model adoption.

\subsubsection{Social Cognitive Theory (SCT)}
The decision to adopt new technologies does not occur in a vacuum. It is influenced by the social milieu within which individuals operate. SCT comes into play here by emphasizing the social and environmental factors that influence individual behaviors~\cite{bandura2014social}. Software engineering, like any profession, has its own culture, norms, and shared beliefs that could significantly shape the adoption of language models. For instance, the prevailing norm or the extent of peer usage could encourage or discourage language model use. Moreover, the self-efficacy of individuals—shaped in part by their social environment—might affect their willingness to engage with such a new tool. SCT, therefore, adds a social layer to our understanding of language model adoption.
\\

The selection of TAM, DOI, and SCT over other potential theories was deliberate and informed by the unique challenges and intricacies of technology adoption in the software engineering domain. While there are numerous theories available that address technology adoption, not all are equally suited to capture the nuances of language model adoption in this specific field. TAM's focus on user perceptions, DOI's emphasis on technological attributes, and SCT's consideration of the social environment together provide a holistic view that other theories might not offer in isolation. Furthermore, the combination of these three theories ensures a multi-dimensional approach, capturing the breadth and depth of factors influencing adoption. Other theories might focus too narrowly on one aspect, potentially overlooking critical influencers. By integrating these three well-established theories, we aimed to achieve a more comprehensive and nuanced understanding, ensuring no significant factor was left unaddressed.

In summary, the triadic theoretical framework of TAM, DOI, and SCT provides a comprehensive lens to examine the adoption of language models in software engineering. By addressing the individual perceptions (TAM), technology characteristics (DOI), and social aspects (SCT), this combined framework provides a well-rounded perspective, ensuring we cover the principal aspects influencing the decision to adopt language models. This choice of theories allows us to glean insightful details that not only offer a rich understanding of the current adoption scenario but also inform strategies to expedite future adoption.


\subsection{Questionnaire Survey Guideline}

Our investigation covers three units of analysis: individual-level factors, technology-level factors, and social-level factors, inspired by the Technology Acceptance Model, the Diffusion of Innovations, and Social Cognitive Theory. We aim to reveal a comprehensive understanding of the acceptance and use of LLM-powered tools in the software engineering context. Here, we detail the final questionnaire survey questions designed to capture these constructs effectively.
As a preliminary step, we conducted a pilot interview with a senior engineering manager from a prominent software company based in Central Europe to ensure the clarity and appropriateness of our questions in April 2023.
To design this investigation, we used the SIGSOFT Empirical Standard for Questionnaire Surveys~\cite{ralph2020empirical}.

Individual-level factors, derived from the Technology Acceptance Model, focus on perceived usefulness, perceived ease of use, and behavioral intention:

\begin{itemize}
    \item \textbf{Perceived Usefulness:}``To what extent do you think using language models increases your efficiency as a software engineer?'' This question is designed to gauge how software engineers perceive the potential productivity gains from using LLMs.
    \item \textbf{Perceived Ease of Use:} ``How easy do you think it is to learn how to use a new language model effectively?'' This question aims to capture the perceived cognitive effort required to learn and adapt to LLMs.
    \item \textbf{Behavioral Intention:} ``How likely are you to use a language model in your work in the next six months? And for which tasks?'' These questions aim to evaluate the intention of software engineers to adopt LLMs in their near-future tasks.
\end{itemize}

Technology-level factors, drawing from the Diffusion of Innovations, include compatibility, relative advantage, and complexity:

\begin{itemize}
    \item \textbf{Compatibility:} ``How is using a language model different from your current software engineering practices?'' This question assesses the perceived fit of LLMs with existing practices and workflows.
    \item \textbf{Relative Advantage:} ``What potential benefits do you think language models offer over your current methods?'' This question helps identify the perceived benefits of LLMs compared to traditional methods.
    \item \textbf{Complexity:} ``What concerns do you have about using a language model in your work?'' This question aims to highlight any perceived barriers or challenges associated with LLM adoption.
\end{itemize}

Social-level factors, grounded in the Social Cognitive Theory, encompass social influence, environmental factors, and self-efficacy:

\begin{itemize}
    \item \textbf{Social Influence:} ``How much do your colleagues or peers influence your decisions to use language models?'' This question examines the impact of social norms and colleagues' opinions on the acceptance of LLMs.
    \item \textbf{Environmental Factors:} ``In your opinion, to what extent do you feel your organization is supportive of adopting language models as a standard technology?'' This question explores the role of organizational support in fostering LLM adoption.
    \item \textbf{Self-Efficacy:} ``How important is it to you to be seen as someone who uses cutting-edge technology in your work?'' This question aims to capture an individual's self-confidence in their ability to use advanced technologies like LLMs effectively.
\end{itemize}

Through these questionnaire survey questions, we strive to understand the complex interplay of individual, technological, and social factors that contribute to the adoption and usage of LLM-powered tools among software engineers.

\subsection{Participants}

The data collection was executed via Prolific Academic~\cite{palan2018prolific}, a well-regarded academic data collection platform often utilized by the software engineering community~\cite{danilova2021you,russo2020gender,russo2020predictors,russo2021daily}. We solicited the input of 100 software engineers, who were asked to answer a series of nine open-ended questions founded on theoretical principles. The compensation for participants (10 minutes) exceeded the US federal minimum wage\footnote{We consistently adhered to the suggested compensation by Prolific. For your reference, participants were paid £9.00 per hour for their time i.e., £1.5 for this investigation.}. The survey was conducted on the Qualtrics platform.

Participants were meticulously chosen through a two-step screening process. Initially, a pre-screening phase was conducted where participants were filtered based on specific self-reported characteristics, including proficiency in computer programming, full-time employment in the software industry, a negative student status, a degree in computer science, and a 100\% approval rate. Following this, a competence screening was performed, as per the methods described by Danilova et al.~\cite{danilova2021you}. This second screening involved assessing participants' knowledge and understanding in key areas, including compilers, programming aid websites, and recursive functions. Furthermore, professionals affirmed their familiarity with Generative AI tools and confirmed their use to a certain extent.

Our data collection methodology complied strictly with the ethical guidelines of the Declaration of Helsinki~\cite{general2014Helsinki}. The Research Ethics Committee at Aalborg University approved this research project in March 2023. All participants were older than 18, gave informed consent prior to participating in the study, and were notified of their right to withdraw their participation at any point. Additionally. the author have completed formal training in research ethics for engineering and behavioral sciences.

In terms of participant demographics, men made up 76\% of the sample, women accounted for 23\%, and non-binary individuals represented 1\%. Geographically, the participants came from various regions: Portugal (23\%), South Africa (15\%), Italy (10\%), United Kingdom (9\%), Poland (9\%), and other countries (34\%).

The professional experience of the participants ranged across various stages in the software industry: 26\% had 0-1 years of experience, 54\% had 2-3 years, 9\% had 4-5 years, 9\% had 6-10 years, and 2\% had over 10 years of experience.

As for their roles in the industry, the majority were software developers or programmers (83\%). This was followed by testers or QA engineers (7\%), data analysts, data engineers, or data scientists (5\%), team leads (3\%), and UX/UI designers (2\%).

\subsection{Analysis of the Qualitative Data}
\label{ssec:QualDatAnal}

Data analysis was implemented within the naturalistic inquiry paradigm~\cite{lincoln1985naturalistic} context, complemented by the constant comparison method~\cite{glaser2017discovery}. The crucial role these strategies play in qualitative data acquisition and examination is significant. This iterative process facilitates initial theory development by identifying patterns and broader dimensions~\cite{gioia1994symbolism}, derived from continual data comparison and analysis, and refining it in accordance with the participants' input \cite{isabella1990evolving}. 

The Thematic Analysis approach was utilized to process the data. Thematic Analysis is a commonly employed method in qualitative research, which involves identifying, analysing, and reporting patterns or themes within the data, while providing a rich, detailed, and complex account of the data~\cite{clarke2015thematic}. The structured methodology proposed by Gioia et al.~\cite{gioia2013seeking} served as the analytical framework. Recent trends within the Management Science community have seen the adoption of this methodology, emphasizing its potential in reinforcing scientific rigour~\cite{linneberg2019coding,grodal2021achieving}. The approach is structured and dedicated to encouraging comprehensive theoretical progression~\cite{gioia2013seeking}. 

The Gioia methodology segments data processing into three stages. The inaugural stage revolves around recognizing \textbf{first-order concepts}, or in-vivo codes~\cite{strauss1990basics}, which align closely with the participants' own words, with minimal researcher-imposed categorization. These codes were then collated into broader themes, a process known as open coding~\cite{vanMaanen1979fact}.

In the subsequent stage, similarities and differences are identified, and emergent themes from these comparisons contribute to the explanation and depiction of the phenomena under investigation. We explored the associations between the concepts to create our high-level themes, employing axial coding.  These are the \textbf{second-order themes}.

The final stage amalgamates similar second-order themes into \textbf{aggregate dimensions}, representing the apex of theoretical contribution. This process was iterative and process-oriented~\cite{locke1996rewriting}, and was perpetuated until theoretical saturation was accomplished~\cite{corbin1990grounded}.

The outcome of this investigation is the \textit{data structure}, which encapsulates first-order terms, second-order themes, and aggregate dimensions for each of the nine theoretical dimensions of our investigation. Notably, the aggregate dimensions were not preconceived categories defined prior to the analysis; rather, they are the end product of a refined and iterative analytical process.

For example, in Table~\ref{tab:TAM-PU}, which details the Data Structure of Perceived Usefulness of LLMs in Software Engineering, a clear progression from first-order concepts to second-order themes and aggregate dimensions is demonstrated. Consider the quote from participant R-15, which is categorized under first-order concepts as ``Automating tasks, reducing time and effort, manual coding, documentation.'' These concepts are then synthesized into the second-order theme of ``Task-specific efficiency improvements,'' indicating a broader category of improvements in efficiency related to specific tasks. Subsequently, this second-order theme is aggregated into the dimension of ``Efficiency Improvement,'' representing a generalized area of impact in software engineering through the use of LLMs. This example illustrates the analytical progression from specific participant quotes to broader thematic categories, demonstrating the methodology of our study.

The presentation of the data structure with their respective second-order themes and first-order concepts are reported in Tables~\ref{tab:TAM-PU}-~\ref{tab:SCT-EF}.

\section{Results}
\label{sec:results}

\subsection{Perceived Usefulness of LLMs in Software Engineering}

The Technology Acceptance Model has been widely used in the study of technology adoption, focusing on two key predictors of acceptance: perceived usefulness and perceived ease of use~\cite{davis1989TAM}. In the context of software engineering, the perceived usefulness of Large Language Models can be examined by evaluating how they contribute to efficiency, productivity, and performance enhancement. Table \ref{tab:TAM-PU} provides a summary of software engineers' perceptions of LLMs in their work.

\subsubsection{Efficiency Improvement}

One of the main perceived benefits of LLMs is their ability to improve efficiency. As shown in Table \ref{tab:TAM-PU}, efficiency improvement is the most frequently mentioned aggregate dimension (55\%). Engineers recognize that LLMs can automate certain tasks, reduce time and effort, and simplify monotonous tasks. For example, R-15 highlighted that LLMs can ``\textit{increase my efficiency by automating certain tasks and reducing the time and effort it takes for manual coding and documentation}.'' This finding aligns with the TAM's emphasis on perceived usefulness, which posits that users will adopt technology if they perceive it to be useful in enhancing their performance~\cite{davis1989TAM}.

\subsubsection{Task-Specific Benefits}

Another aspect of perceived usefulness is the task-specific benefits LLMs provide, such as debugging, learning new features, and generating code snippets. As R-30 mentioned, LLMs have significantly increased their efficiency by helping them ``\textit{debug code faster, learn about new features without scanning the whole documentation, and providing useful code snippets for work}.'' This category represents 26\% of the aggregate dimensions and supports the notion that perceived usefulness is an important predictor of LLM adoption~\cite{venkatesh2003user}.

\subsubsection{Complementary Tool}

LLMs are also viewed as a complementary tool to human expertise and judgment. R-17 pointed out that ``\textit{language models have the potential to enhance productivity and efficiency in software engineering, but they should be used as a tool alongside human expertise and judgment}.'' This perception highlights the importance of balancing the benefits of LLMs with the need for human oversight, an aspect that may influence the overall perceived usefulness of the technology.

\subsubsection{Limited Applicability and Quality Concerns}

While many respondents reported positive perceptions of LLMs, some expressed concerns about their limited applicability (15\%) and quality concerns (9\%). For instance, R-21 mentioned that LLMs are ``\textit{nice for generic tasks, but the models have zero knowledge about our internal APIs so they're really hard to apply}.'' R55 also noted that while LLMs may help, ``\textit{you have to review and understand the code anyway. So I don't know if it makes you more efficient}.'' These concerns suggest that while LLMs can offer benefits in certain situations, their usefulness may be limited by the need for review and adaptation to specific contexts. This finding is consistent with the TAM literature, which highlights that the perceived usefulness of a technology is not only determined by its benefits but also by its limitations~\cite{venkatesh2003user}.
\\

The perceived usefulness of LLMs in software engineering, as reflected in the efficiency improvement, task-specific benefits, and complementary nature of the technology, supports the potential for widespread adoption. However, the concerns related to limited applicability and quality highlight the importance of addressing these limitations to enhance the perceived usefulness and, consequently, the acceptance of LLMs. This analysis aligns with the TAM framework, which emphasizes that perceived usefulness is a critical determinant of technology acceptance~\cite{davis1989TAM}.

\begin{table}[!ht]
\centering
\tiny
\caption{Data Structure of Perceived Usefulness of LLMs in Software Engineering}
\begin{tabular}{p{0.7cm}p{5.5cm}p{1.7cm}p{1.7cm}p{1.7cm}p{0.5cm}}
\toprule
\textbf{ID} & \textbf{Quote} & \textbf{1st Order Concepts} & \textbf{2nd Order Themes} & \textbf{Aggregate Dimensions} & \textbf{(\%)} \\
\midrule
R-15 & The can greatly increase my efficiently by automating certain tasks and reducing the time and effort it take for manual coding and documentation & Automating tasks, reducing time and effort, manual coding, documentation & Task-specific efficiency improvements & Efficiency Improvement & 55 \\
R-28 & it increases considerably my efficiency specially in simple tasks & Increased efficiency, simple tasks & Task-specific efficiency improvements & Efficiency Improvement & 55 \\
R-52 & They help only in monotonous and simple tasks (defining constructors and writing user input validating systems for example). & Monotonous tasks, simple tasks & Task-specific efficiency improvements & Efficiency Improvement & 55 \\
R-67 & I save 10\% - 20\% of time & Time savings & Time savings & Time Savings & 24 \\
R-30 & It certainly helps a lot, these last few days that I've been particularly using ChatGPT (with GPT 3-5), my efficiency has gone up by quite a bit. It helps me debug code faster, learn about new features without scanning the whole documentation, and providing me useful code snippets for my work. & Increased efficiency, debugging, learning new features, code snippets & Task-specific benefits, learning enhancement & Task-Specific Benefits & 26 \\
R-17 & I think language models can increase the efficiency of software engineers by automating certain tasks, such as code generation, testing, and documentation. Additionally, language models can help with data analysis and decision-making, allowing engineers to make informed choices based on large datasets. Overall, language models have the potential to enhance productivity and efficiency in software engineering, but they should be used as a tool alongside human expertise and judgement. & Automating tasks, data analysis, decision-making, human expertise, judgement & Complementary tool, efficiency improvement & Complementary Tool & 18 \\
R-21 & Very slightly. It's nice for generic tasks, but the models have zero knowledge about our internal APIs so they're really hard to apply. & Limited applicability, internal APIs, generic tasks & Limited applicability & Limited Applicability & 15 \\
R-41 & I'm much more efficient using a language model as it have been helping me to understand the company's code much faster. & Increased efficiency, understanding company code, faster learning & Learning enhancement & Learning Enhancement & 12 \\
R-76 & While most of the time, language models get small things wrong, thereby requiring extra time for checking their output, the time they save by doing especially the boring parts of coding for you definitely outweighs this in my opinion. I would say using copilot for example has increased my coding efficiency by about 30\%. & Time savings, checking output, increased efficiency & Time savings, quality concerns & Time Savings, Quality Concerns & 24,9 \\
R-55 & I think it helps but then you have to review and understand the code anyway. So I don't know if it makes you more efficient. & Code review, understanding, efficiency concerns & Quality concerns & Quality Concerns & 9 \\
\bottomrule
\end{tabular}
\label{tab:TAM-PU}
\end{table}

\subsection{Perceived Ease of Use of LLMs in Software Engineering}

In this subsection, we present the results of our qualitative analysis of the questionnaire survey statements, highlighting the key factors that influence the perceived ease of use of Large Language Models in software engineering. Our analysis draws upon the Technology Acceptance Model framework~\cite{davis1989TAM}, which posits that the perceived ease of use and perceived usefulness are essential determinants of technology adoption. We have identified several aggregate dimensions that explain the perceived ease of use of LLMs and present them in separate subsubsections, providing empirical evidence from the questionnaire survey statements (Table \ref{tab:TAM-PEOU}).

\subsubsection{Learning Process}

Our analysis reveals that the learning process is a crucial factor influencing the perceived ease of use of LLMs in software engineering. As shown in Table \ref{tab:TAM-PEOU}, R-54 reported that ``\textit{once I got familiar with the technology, it became much easier to use}.'' This finding aligns with the Technology Acceptance Model proposed by~\cite{davis1989TAM}, which suggests that the perceived ease of use of a technology is directly related to its adoption. Moreover, prior research has emphasized the role of learning in the adoption of new technologies~\cite{venkatesh2000theoretical}. In this context, the learning curve associated with LLMs appears to be an essential determinant of their perceived ease of use.

\subsubsection{Prior Experience}

The questionnaire survey data also underscore the importance of prior experience in shaping the perceived ease of use of LLMs. For example, R-20 stated that ``\textit{I think it is easy when you know the different concepts}.'' This observation is consistent with the literature on technology adoption, which suggests that individuals with prior experience in related technologies are more likely to perceive a new technology as easy to use~\cite{compeau1995computer}. In the case of LLMs, having a background in programming languages or natural language processing (NLP) could facilitate their adoption in software engineering.

\subsubsection{Individual Differences}

Another key theme that emerged from our analysis is the role of individual differences in shaping the perceived ease of use of LLMs. As R-9 noted, ``\textit{it depends on the person and how they are used to work}.'' This finding supports the notion that individual characteristics, such as cognitive style and personal innovativeness, can influence the perceived ease of use of a technology~\cite{segars1993re}. In the context of LLMs, the extent to which software engineers perceive them as easy to use may depend on their unique preferences, learning styles, and problem-solving approaches.

\subsubsection{Intuitiveness and User Interface}

The intuitiveness of LLMs and their user interface design also emerged as important factors in our analysis. For instance, R-89 mentioned that ``\textit{they were pretty much made for ease of use by the average consumer}.'' This observation aligns with the work of~\cite{nov2008users}, who argued that a well-designed user interface can significantly enhance the perceived ease of use of a technology. In the case of LLMs, an intuitive and user-friendly interface could facilitate their adoption among software engineers.

\subsubsection{Task Complexity}

Finally, the complexity of the tasks that LLMs are used for in software engineering appears to influence their perceived ease of use. As R-53 noted, ``\textit{the difficulty to learn how to use them effectively can vary as it depends on how you're using it, but for the most part, it ranges from not too hard to very hard}.'' This finding is consistent with the Task-Technology Fit model~\cite{goodhue1995task}, which posits that thefit between the technology and the task it is intended for affects the technology's perceived ease of use and, ultimately, its adoption. In the context of LLMs, it seems that software engineers may find them easier to use for certain tasks, while others might require a higher level of expertise and knowledge.
\\

In summary, our analysis identified several aggregate dimensions that explain the perceived ease of use of LLMs in software engineering, including the learning process, prior experience, individual differences, intuitiveness and user interface, and task complexity. These factors provide a nuanced understanding of the adoption of LLMs in software engineering and their connection to the theoretical framework of the Technology Acceptance Model. By incorporating the empirical evidence from the questionnaire survey statements and drawing on relevant literature, our findings contribute to the ongoing conversation about the role of LLMs in software engineering and the factors that influence their adoption.

\begin{table}[!ht]
\centering
\tiny
\caption{Data Structure of Perceived Ease of Use of LLMs in Software Engineering}
\begin{tabular}{p{0.7cm}p{5.5cm}p{1.7cm}p{1.7cm}p{1.7cm}p{0.5cm}}
\toprule
\textbf{ID} & \textbf{Quote} & \textbf{1st Order Concepts} & \textbf{2nd Order Themes} & \textbf{Aggregate Dimensions} & \textbf{(\%)} \\
\midrule
R-46 & easy as you practice more & Practice, Improvement & Practice and improvement & Learning Process & 54 \\ 
R-4 & it takes some time and dedication & Time, Dedication & Learning curve & Learning Process & 54 \\ 
R-96 & However, for more advanced tasks, it can take some time to word your questions correctly & Advanced tasks, Time & Learning curve & Learning Process & 54 \\ 
R-30 & In the first week of using it, you will be already increasing your efficiency & Efficiency, Time & Practice and improvement & Learning Process & 54 \\ 
R-33 & To be more efficient, it requires to learn the specific prompts that give you an exact answer you expect & Efficiency, Specific prompts & Practice and improvement & Learning Process & 54 \\ 
R-17 & depends on the complexity and capabilities of the model, as well as the user's prior experience and knowledge & Complexity, User experience, Prior knowledge & Individual factors & Individual Background & 26 \\ 
R-98 & It is easy when you know another language that is similar to it & Prior knowledge, Similar language & Individual factors & Individual Background & 26 \\ 
R-5 & Extremely easy & Ease of use & Perceived ease of use & Perceived Ease of Adoption & 13 \\ 
R-52 & Easy to use, hard to master the prompts & Ease of use, Mastery & Mastery & Perceived Ease of Adoption & 13 \\ \
R-74 & The difficult part is to understand if the response given is correct and related to what someone needs & Response quality, Relation to needs & Perceived difficulty & Model Effectiveness & 6 \\ 
\bottomrule
\end{tabular}
\label{tab:TAM-PEOU}
\end{table}

\subsection{Behavioral Intention of LLMs in Software Engineering}
The Technology Acceptance Model has been widely used to understand the factors influencing the adoption of new technologies in various contexts, such as software engineering~\cite{davis1989user,venkatesh2000theoretical}. According to the TAM, behavioral intention, which reflects the likelihood of an individual to use a specific technology, is influenced by two main factors: perceived usefulness and perceived ease of use~\cite{davis1989user}. In this subsection, we explore the behavioral intention of software engineers in relation to the adoption of Large Language Models, focusing on the aggregate dimensions emerged from the performed analysis (Table \ref{table:TAM-BI}).

\subsubsection{Code Improvement and Maintenance}
A significant portion of software engineers indicated their intention to use LLMs to improve and maintain their codebase. This finding aligns with the TAM's concept of perceived usefulness, as using LLMs for code refactoring, adherence to design patterns, and implementation of SOLID principles can enhance software quality and maintainability~\cite{peruma2022refactor}. For instance, R-8 mentioned, ``\textit{I am considering purchasing a ChatGPT-4 subscription, mainly to refactor (legacy) code or to make it adhere to certain design patterns. It could also help refactoring code to make it more SOLID.}'' This quote highlights the potential value of LLMs in addressing common software engineering challenges.

\subsubsection{Efficiency and Automation}
LLMs were perceived to be useful for automating repetitive tasks and increasing efficiency. This perception corresponds to both perceived usefulness and perceived ease of use in the TAM, as task automation can lead to time savings and streamline the development process~\cite{venkatesh2000theoretical}. R-35 stated, ``\textit{I believe I may start using a language model more and more, especially to automate tasks which can be performed by a language model and which take a significant amount of time.}'' The adoption of LLMs for task automation can potentially improve software engineers' productivity.

\subsubsection{Learning and Problem Solving}
The use of LLMs for learning and problem solving was another theme identified, which is in line with the TAM's perceived usefulness. Software engineers expressed the intention to use LLMs for tasks such as finding documentation, clarifying confusing code, and seeking information on programming-related questions. As R-25 stated, ``\textit{Very likely. Mostly to find documentation for libraries, refactor and clarify confusing code.}'' LLMs can serve as a valuable learning and problem-solving tool for software engineers, supporting continuous professional development.

\subsubsection{Specialized Applications}
Respondents mentioned the potential use of LLMs for specific tasks, such as writing basic functionalities, defining tasks, and composing emails. This theme is related to the perceived usefulness of LLMs in addressing particular software engineering needs. R-62, for example, mentioned, ``\textit{Very likely. For writing basic functionalities, defining tasks, for emails.}'' The adoption of LLMs for specialized applications can provide targeted benefits to software engineers in their daily work.

\subsubsection{Adoption Barriers and Concerns}
Despite the potential benefits of LLMs, some software engineers expressed concerns and barriers to adoption, such as cost, dependency on third-party services, and potential ethical issues. These concerns align with the TAM's concept of perceived ease of use, as they can hinder the adoption of LLMs~\cite{venkatesh2000theoretical}. For instance, R-60 stated, ``\textit{The cost and dependency on a third-party service might be a concern.}'' Understanding and addressing these concerns is essential for promoting LLM adoption in software engineering.
\\

In conclusion, our analysis reveals several factors that influence the behavioral intention of software engineers to adopt LLMs, aligning with the theoretical framework of the TAM. LLMs are perceived as useful for code improvement and maintenance, efficiency and automation, learning and problem-solving, and specialized applications, while some adoption barriers and concerns persist. By understanding these factors, we can better support the integration of LLMs into software engineering practices and promote their adoption to enhance productivity and software quality.

\begin{table}[!ht]
\centering
\tiny
\caption{Data Structure of Behavioral Intention of LLMs in Software Engineering}
\begin{tabular}{p{0.7cm}p{5.5cm}p{1.7cm}p{1.7cm}p{1.7cm}p{0.5cm}}
\toprule
\textbf{ID} & \textbf{Quote} & \textbf{1st Order Concepts} & \textbf{2nd Order Themes} & \textbf{Aggregate Dimensions} & \textbf{(\%)} \\
\midrule
R-8 & Very likely. I am considering purchasing a ChatGPT-4 subscription, mainly to refactor (legacy) code or to make it adhere to certain design patterns. It could also help refactoring code to make it more SOLID. & Refactor code, Design patterns, SOLID principles & Code generation and refactoring & Code Improvement and Maintenance & 42 \\

R-35 & I believe I may start using a language more and more, especially to automate tasks which can be performed by a language model and which take a significant amount of time. & Automate tasks, Time-saving & Task automation and optimization & Efficiency and Automation & 35 \\

R-25 & Very likely. Mostly to find documentation for libraries, refactor and clarify confusing code. & Find documentation, Refactor code, Clarify confusing code & Information seeking and learning & Learning and Problem Solving & 28 \\

R-7 & Tried it out with some basic programming related questions already. & Basic programming questions & Information seeking and learning & Learning and Problem Solving & 28 \\

R-62 & Very likely. For writing basic functionalities, defining tasks, for emails & Writing basic functionalities, Defining tasks, Emails & Task-specific use cases & Specialized Applications & 20 \\

R-17 & Very likely in language translation, text summarization and text generation & Language translation, Text summarization, Text generation & Natural language processing tasks & NLP and Content Generation & 18 \\

R-46 & Writing automated tests & Writing automated tests & Testing and code validation & Quality Assurance and Validation & 15 \\

R-69 & Not likely & Non-adoption & Uncertainty or non-adoption & Adoption Barriers and Concerns & 12 \\

R-60 & The cost and dependency on a third-party service might be a concern. & Cost, Dependency on third-party service & Adoption barriers & Adoption Barriers and Concerns & 12 \\
\bottomrule
\end{tabular}
\label{table:TAM-BI}
\end{table}

\subsection{Compatibility of LLMs in Software Engineering}

The compatibility of Large Language Models in software engineering is a crucial factor in understanding their adoption and impact on software development practices. Compatibility refers to the degree to which an innovation is perceived as being consistent with existing values, past experiences, and the needs of potential adopters~\cite{rogers2010diffusion}. This subsection presents a thematic analysis of the compatibility of LLMs in software engineering based on the responses of software engineers, which are summarized in Table~\ref{tab:DOI-Compatibility}. We have identified four aggregate dimensions, as detailed in the following sub-subsections: Improved Efficiency, Assistance and Support, Similarity to Current Practices, and Adaptation and Learning.

\subsubsection{Improved Efficiency}

Improved Efficiency was the most frequently occurring theme in the data, with 39\% of the responses reflecting this aspect of compatibility. The use of LLMs in software engineering tasks is perceived to speed up the development process, automate mundane tasks, and ultimately improve overall efficiency. One respondent (R-19) highlighted that LLMs ``\textit{can be used to automate certain tasks in software engineering},'' thus reducing the time and effort spent on repetitive tasks. Similarly, R-35 noted that using a language model ``\textit{will speed up the tasks of going to look for specific pieces of code from multiple websites}.'' These findings align with the literature, where compatibility is a key factor for technology adoption~\cite{tornatzky1982innovation}.

\subsubsection{Assistance and Support}

Assistance and Support emerged as the second most frequent theme in the data, representing 28\% of the responses. Respondents highlighted the value of LLMs in providing help and support, particularly in situations where traditional search methods fail to deliver desired results. R-1 mentioned that LLMs can provide an ``\textit{extra hand and assistance for things I don't know and can't find with a traditional search}.'' This demonstrates that LLMs' ability to offer contextually relevant and targeted assistance is seen as an important aspect of their compatibility with software engineering practices.

\subsubsection{Similarity to Current Practices}

Similarity to Current Practices was reported by 16\% of the respondents. This theme suggests that the adoption of LLMs in software engineering is facilitated by their perceived similarities to existing tools and practices. R-15 noted that there is ``\textit{not much of a difference as the language model is just used to assist my current software engineering practices}.'' R-5 similarly stated that using LLMs is ``\textit{like Googling but I don't need to filter as much information}.'' The perceived similarity to current practices can influence the adoption of LLMs as it reduces the barriers to their integration into existing workflows~\cite{tornatzky1982innovation,rogers2010diffusion}.

\subsubsection{Adaptation and Learning}

Adaptation and Learning was reported by 11\% of the respondents. This theme highlights the importance of learning and adapting to new tools and techniques in the software engineering domain. R-46 expressed that using LLMs is ``\textit{something new: you have to learn how to use it}.'' Similarly, R-87 mentioned that using a language model ``\textit{represents a new paradigm for me}.'' This theme indicates that the compatibility of LLMs in software engineering can be enhanced by promoting learning and adaptation among software engineers. The adoption of LLMs can thus be facilitated by providing resources and training to help engineers understand and integrate these tools into their daily work.
\\

In conclusion, this subsection has explored the compatibility of LLMs in software engineering by analyzing the aggregate dimensions derived from the responses of software engineers. These dimensions—Improved Efficiency, Assistance and Support, Similarity to Current Practices, and Adaptation and Learning—provide a comprehensive understanding of the factors that influence the compatibility of LLMs in software engineering. By linking these dimensions to the Diffusion of Innovation theory~\cite{rogers2010diffusion}, this analysis offers valuable insights into the factors that contribute to the adoption and integration of LLMs into software engineering practices. This understanding can help inform the development of LLMs that are more compatible with existing workflows and practices, ultimately leading to more widespread adoption and use in the software engineering domain.

\begin{table}[!ht]
\centering
\tiny
\caption{Data Structure of Compatibility of LLMs in Software Engineering}
\begin{tabular}{p{0.7cm}p{5.5cm}p{1.7cm}p{1.7cm}p{1.7cm}p{0.5cm}}
\toprule
\textbf{ID} & \textbf{Quote} & \textbf{1st Order Concepts} & \textbf{2nd Order Themes} & \textbf{Aggregate Dimensions} & \textbf{(\%)} \\
\midrule
R-19       & Language models can be used to automate certain tasks in software engineering...                             & Automate tasks, Speed up process & Improved software engineering tasks & Improved Efficiency         & 39      \\ 
R-27       & It shifts the burden of research from me to the algorithm.                                                    & Shift burden, Algorithm           & Improved software engineering tasks & Improved Efficiency         & 39      \\ 
R-35       & A language model will speed up the tasks of going to look for specific pieces of code from multiple websites. & Speed up tasks, Code search       & Improved software engineering tasks & Improved Efficiency         & 39      \\ 
R-1        & It gives an extra hand and can provide assistance for things I don't know and can't find with a traditional search. & Extra hand, Assistance           & Providing help and support          & Assistance and Support      & 28      \\ 
R-15       & There is not much of a difference as the language model is just used to assist my current software engineering practices. & Similarity, Assistance           & Similarity to current practices     & Similarity to Current Practices & 16      \\ 
R-5        & Not that different, it's like Googling but I don't need to filter as much information.                       & Similarity, Less filtering        & Similarity to current practices     & Similarity to Current Practices & 16      \\ 
R-46       & It's something new: you have to learn how to use it.                                                          & Learning, Adaptation              & Need for adaptation and learning    & Adaptation and Learning     & 11      \\ 
R-87       & Using a language model represents a new paradigm for me...                                                   & New paradigm, Adaptation          & Need for adaptation and learning    & Adaptation and Learning     & 11      \\ 
R-99       & Far more interactive and personalized compared to normal google searches.                                    & Interactive, Personalized         & Personalized and interactive use    & Personalization and Interaction & 10      \\ 
R-58       & It offers a more specific and tailored response compiled from a lot of content available online...            & Specific, Tailored response       & Personalized and interactive use    & Personalization and Interaction & 10      \\ 
\bottomrule
\end{tabular}
\label{tab:DOI-Compatibility}
\end{table}

\subsection{Complexity of LLMs in Software Engineering}
The complexity of adopting Large Language Models in software engineering is a critical aspect of understanding their diffusion and impact on the industry. According to the Diffusion of Innovation theory, the complexity of an innovation influences its adoption rate, with more complex innovations facing a slower adoption~\cite{rogers2010diffusion}. The thematic analysis of the questionnaire survey data (Table \ref{tab:DOI-Complexity}) reveals several aggregate dimensions that contribute to the perceived complexity of LLMs in software engineering. In this subsection, we discuss each of these dimensions and their implications for the adoption of LLMs, linking them to the relevant literature and the Diffusion of Innovation theory.

\subsubsection{Job Security Concerns}
The fear of job loss and skill devaluation emerged as a significant concern among respondents (25\% frequency). Respondent R-15 stated, ``\textit{I'm concerned that it can automate a lot of tasks and make most of my work obsolete}.'' This perspective aligns with research on the potential disruptive effects of AI and automation on the job market~\cite{frey2017future}. According to the Diffusion of Innovation theory, innovations that are perceived to threaten job security are likely to face resistance~\cite{rogers2010diffusion}. To mitigate this challenge, organizations should communicate the benefits of LLMs and focus on upskilling and reskilling employees~\cite{tenakwah2021employees,bessen2019automation}.

\subsubsection{Dependence and Complacency}
Some respondents (16\% frequency) expressed concerns about junior programmers relying too much on LLMs, leading to a decline in code understanding and increased reliance on these models. Respondent R-34 explained, ``\textit{My concern is other junior programmers using it without understanding the code and causing bugs (more work for me)}.'' This challenge can be addressed by promoting the responsible use of LLMs and ensuring that programmers have a strong foundation in coding concepts.

\subsubsection{Data Security and Privacy}
Data security and privacy concerns were identified by 15\% of respondents. They expressed concerns about LLMs being trained on sensitive data, potentially leading to privacy breaches. Respondent R-17 mentioned, ``\textit{In terms of privacy, as language models can be trained on sensitive or personal data, such as emails, messages, or documents. This may raise privacy and data protection concerns}.'' To address this issue, developers should ensure that LLMs are trained on secure, anonymized datasets and that privacy regulations are followed.

\subsubsection{Quality and Accuracy of Generated Code}
Respondents also raised concerns about the quality and accuracy of code generated by LLMs (13\% frequency). Respondent R-49 remarked, ``\textit{Language models aren't perfect, so I would be afraid that they would cause errors}.'' Ensuring the reliability and accuracy of generated code is essential for LLM adoption~\cite{ray2014large}. To address this issue, developers should establish best practices for code review and validation, as well as invest in improving the models' performance~\cite{wei2019code}.

\subsubsection{Ethical and Legal Considerations}
Ethical and legal considerations, such as authorship and intellectual property rights, were identified by 8\% of respondents. Respondent R-28 simply stated, ``\textit{Author rights are tricky to attribute}.'' Organizations should consider the ethical implications of using LLMs and establish guidelines for their use to ensure compliance with existing laws and regulations~\cite{mittelstadt2016ethics}.

\subsubsection{Bias and Explainability}
Some respondents (7\% frequency) highlighted concerns about the potential biases in outputs and the lack of explainability in their decision-making processes. Respondent R-62 expressed, ``\textit{The biases in the model can have unintended consequences}.'' It is essential to address these issues to ensure the responsible use of LLMs in software engineering. Organizations can invest in research to reduce biases and improve the explainability of LLMs to enhance their trustworthiness and adoption~\cite{arrieta2020explainable}.

\subsubsection{Integration and Compatibility}
Integration and compatibility issues were mentioned by 6\% of respondents, who expressed concerns about the ability of LLMs to work seamlessly with existing software development tools and practices. Respondent R-21 stated, ``\textit{Integrating the model into the current workflow might be challenging}.'' To facilitate the adoption of LLMs, developers should ensure that these models are compatible with existing tools and can be easily integrated into the software development process.
\\

In conclusion, the complexity of LLM adoption in software engineering is multifaceted, encompassing various concerns, such as job security, dependence, data security, code quality, ethical issues, bias, and integration challenges. By addressing these concerns, organizations can facilitate the adoption of LLMs and leverage their potential benefits in software engineering. This discussion, grounded in the thematic analysis of the questionnaire survey data and the Diffusion of Innovation theory, contributes to the understanding of the factors that affect LLM adoption in the software engineering context.

\begin{table}[!ht]
\centering
\tiny
\caption{Data Structure of Complexity of LLMs in Software Engineering}
\begin{tabular}{p{0.7cm}p{5.5cm}p{1.7cm}p{1.7cm}p{1.7cm}p{0.5cm}}
\toprule
\textbf{ID} & \textbf{Quote} & \textbf{1st Order Concepts} & \textbf{2nd Order Themes} & \textbf{Aggregate Dimensions} & \textbf{(\%)} \\
\midrule
R-15 & I'm concerned that it can automate a lot of tasks and make most of my work obsolete. & Automation, task replacement, obsolescence & Job loss, skill devaluation & Job security concerns & 25 \\ 
R-34 & My concern is other junior programmers using it without understanding the code and causing bugs (more work for me). & Junior programmers, lack of understanding, bugs & Decline in code understanding, reliance on language models & Dependence and complacency & 16 \\ 
R-17 & In terms of privacy, as language models can be trained on sensitive or personal data, such as emails, messages, or documents. This may raise privacy and data protection concerns, especially if the language model is used in a cloud-based environment or shared with third parties. & Privacy, data protection, sensitive data & Data confidentiality, exposure risk & Data security and privacy & 15 \\ 
R-49 & Language models aren't perfect, so I would be afraid that they would cause errors. & Language model imperfection, errors & Incorrect or poorly maintained code & Quality and accuracy of generated code & 13 \\ 
R-28 & Author rights are tricky to attribute & Authorship, rights & Legal concerns & Ethical and legal considerations & 8 \\ 
R-87 & Concerns I have about using language models in my work include issues of bias, explainability, and potential security vulnerabilities. & Bias, explainability, security vulnerabilities & Trust in generated results, interpretability & Bias and explainability & 7 \\ 
R-6 & Might not be compatible with the systems at my workplace. & Compatibility, systems & Integration challenges & Compatibility and integration & 5 \\ 
R-5 & My programming skills might get ``rusty`` if I rely too much on it, like with autocorrect, I feel my grammar is now worse because my phone understands even incomplete words and I can't remember the proper way of writing some words because of this. & Relying on language models, skill deterioration, autocorrect & Personal skill decline & Skill deterioration & 3 \\ 
\bottomrule
\end{tabular}
\label{tab:DOI-Complexity}
\end{table}

\subsection{Relative Advantage of LLMs in Software Engineering}
The relative advantage of Large Language Models in software engineering is a key factor in their adoption, as postulated by the Diffusion of Innovation theory~\cite{rogers2010diffusion}. Our qualitative data analysis, based on Gioia's Methodology, highlights the various aspects of LLMs that contribute to their perceived benefits over current methods. The following sub-subsections present the Aggregate Dimensions derived from our analysis and discuss their implications in relation to the relative advantage construct (Table \ref{tab:DOI-RA}).

\subsubsection{Time Efficiency}
A recurring theme in our analysis was the time efficiency provided by LLMs, as reported by 42\% of respondents. Respondents appreciated the ability of LLMs to quickly complete tasks, search for relevant information, and provide solutions to coding issues. For example, R-25 noted that LLMs significantly reduced time spent searching for information: \textit{``I believe the best thing is the time spent searching for certain things is way lower than before}.'' This time efficiency can be attributed to the natural language processing capabilities of LLMs, which enable users to communicate their needs more effectively and rapidly obtain tailored solutions~\cite{brown2020language}.

\subsubsection{Code Quality}
Improvements in code quality emerged as another key advantage of LLMs, with 14\% of respondents highlighting the positive impact on their work. Respondents reported that LLMs provided clearer, more understandable code, which reduced errors and improved overall code robustness. R-37 stated: \textit{``It generates a simpler and more understandable code, working in a more organized way and reducing errors}.'' This improvement in code quality is a result of LLMs' ability to analyze and learn from vast amounts of source code, allowing them to provide optimal solutions based on best practices~\cite{devlin2019bert}.

\subsubsection{User Experience}
LLMs were noted to enhance the overall user experience, with 11\% of respondents mentioning the ease of use and communication with the models. R-67 commented on the human-like interaction: \textit{``Language models are easier to use and faster because you can send messages like for human}.'' The improved user experience can be attributed to the natural language understanding capabilities of LLMs, which allow them to interpret and respond to user inputs more effectively than traditional methods~\cite{radford2019language}.

\subsubsection{Learning and Skill Development}
LLMs were also found to facilitate learning and skill development, as reported by 9\% of respondents. Respondents appreciated the ability of LLMs to simplify the learning process and reduce the time spent on mastering technical concepts. R-23 explained: \textit{``I think with language models you don't have to spend that much time learning `technical' things}.'' This can be linked to the contextual understanding and knowledge retention capabilities of LLMs, which allow them to provide tailored guidance and support for users with varying levels of expertise.

\subsubsection{Customization and Personalization}
Finally, customization and personalization were highlighted as advantages by 8\% of respondents. LLMs were praised for their ability to provide more digestible information and adapt responses to user preferences. R-94 described the flexibility of LLMs: \textit{``It gives more digested information, and we can 'mold' the information how we want (e.g., asking the language model to respond using short sentences, or to explain in detail certain topics, etc)}.'' This aspect of LLMs can be attributed to their capacity for understanding context and user preferences, allowing them to generate more relevant and personalized responses~\cite{openai2023gpt4}.
\\

In summary, our thematic analysis revealed several key aspects of LLMs that contribute to their perceived relative advantage in software engineering, including time efficiency, code quality, user experience, learning and skill development, and customization and personalization. These findings align with the Diffusion of Innovation theory, suggesting that the adoption of LLMs in software engineering can be facilitated by their ability to provide clear benefits over existing methods~\cite{rogers2010diffusion}. Moreover, our results highlight the potential of LLMs to revolutionize the field of software engineering by streamlining processes, enhancing user experience, and fostering continuous learning and improvement.

\begin{table}[!ht]
\centering
\tiny
\caption{Data Structure of Relative Advantage of LLMs in Software Engineering}
\begin{tabular}{p{0.7cm}p{5.5cm}p{1.7cm}p{1.7cm}p{1.7cm}p{0.5cm}}
\toprule
\textbf{ID} & \textbf{Quote} & \textbf{1st Order Concepts} & \textbf{2nd Order Themes} & \textbf{Aggregate Dimensions} & \textbf{(\%)} \\
\midrule
R-25 & I believe the best thing is the time spent searching for certain things is way lower than before. & Time-saving search & Task Completion & Time Efficiency & 42 \\ 
R-37 & It generates a simpler and more understandable code, working in a more organized way and reducing errors. & Understandable code, Reduced errors & Code Simplicity \& Robustness & Code Quality & 14 \\ 
R-67 & Language models are easier to use and faster because you can send messages like for human & Easier to use, Faster & Ease of Communication & User Experience & 11 \\ 
R-23 & I think with language models you don't have to spend that much time learning 'technical' things & Less time learning & Simplified Learning & Learning and Skill Development & 9 \\ 
R-64 & It can help me jump over boilerplate code and snippets I typed hundreds of times before. It can also help in learning new languages' syntax. & Boilerplate code, Learning new languages & Code Generation \& Automation & Automation \& Adaptability & 8 \\ 
R-35 & I believe it might be more efficient and save some time. & Efficient, Time-saving & Innovative Solutions & Creativity \& Innovation & 3 \\ 
R-11 & I think it can bring clear points of view to the table that weren't take into consideration & New points of view & Diverse Perspectives & Insights \& Decision-making & 4 \\ 
R-68 & More effective and faster problem resolution & Faster problem resolution & Effective Troubleshooting & Problem-solving & 6 \\ 
R-53 & Language models (paired programming) can help make fewer mistakes and overall increase efficiency. & Fewer mistakes, Increased efficiency & Collaborative Assistance & Teamwork \& Collaboration & 5 \\ 
R-94 & It gives more digested information, and we can ``mold`` the information how we want (e.g. asking the language model to respond using short setences, or to explain in detail certain topics, etc) & Digestible information, Customization & Flexible Responses & Customization \& Personalization & 8 \\ 
\bottomrule
\end{tabular}
\label{tab:DOI-RA}
\end{table}

\subsection{Social Influence of LLMs in Software Engineering}
The adoption of Large Language Models in software engineering is influenced by various factors. One key factor is the social influence from peers and colleagues, as suggested by the Social Cognitive Theory~\cite{bandura2014social}. The current analysis aims to provide a rationale on how the data explains the ``social influence'' in relation to the adoption of LLMs in software engineering and how it links to the theoretical framework of the Social Cognitive Theory. Our analysis (Table \ref{tab:SCT-SI}) revealed four aggregate dimensions representing the range of social influences in the adoption of LLMs: No Influence, Low Influence, Moderate Influence, and High Influence.

\subsubsection{No Influence}
Our analysis revealed that 29\% of the respondents reported no influence from their colleagues or peers on their decision to use LLMs (e.g., R-66, R-24, R-58). This finding suggests that a considerable proportion of software engineers make independent decisions about whether to adopt LLMs. This aligns with the literature on individual agency and self-efficacy in the Social Cognitive Theory~\cite{bandura1997SelfEfficacy}. These software engineers may rely on their own evaluation of the technology and personal preferences, rather than the opinions or experiences of their colleagues.

\subsubsection{Low Influence}
Low influence was reported by 24\% of the respondents (e.g., R-45, R-99). This indicates that some software engineers may be slightly influenced by their peers, but ultimately retain a high degree of autonomy in their decision-making. This finding suggests that while social influence may play a role in the adoption of LLMs, individual factors, such as personal interest and perceived utility, may also significantly contribute to the decision-making process~\cite{ajzen2020theory}.

\subsubsection{Moderate Influence}
Our analysis revealed that 21\% of the respondents reported moderate influence from their peers and colleagues (e.g., R-9, R-82). This suggests that a significant proportion of software engineers value the input and feedback of their peers when deciding whether to adopt LLMs. This finding is consistent with the Social Cognitive Theory, which emphasizes the role of observational learning and vicarious experiences in shaping individual behavior~\cite{bandura2014social}. In the context of LLM adoption, this moderate influence may result from a combination of individual factors and the experiences of colleagues.

\subsubsection{High Influence}
Finally, 26\% of the respondents reported high influence from their peers and colleagues (e.g., R-97, R-79, R-27). This finding suggests that a considerable proportion of software engineers are strongly influenced by the collective use and enthusiasm for LLMs within their professional circles. This result aligns with the Social Cognitive Theory's focus on the reciprocal interactions between individuals and their social environment, as well as the literature on technology adoption in organizations~\cite{rogers2010diffusion}. In this case, the high level of influence may stem from the perceived benefits of LLMs, collaboration, and shared enthusiasm for exploring the technology's possibilities.
\\

In summary, our analysis revealed a diverse range of social influence levels in the adoption of LLMs in software engineering. While some software engineers reported no influence from their colleagues or peers, others indicated low, moderate, or high levels of influence. These findings highlight the complex interplay between individual factors, such as self-efficacy and personal interest, and social influences, as posited by the Social Cognitive Theory. The understanding of these various levels of social influence can inform future research on the adoption of LLMs and other emerging technologies in software engineering, as well as organizational strategies for encouraging their appropriate use.

\begin{table}[!ht]
\centering
\tiny
\caption{Data Structure of Social Influence of LLMs in Software Engineering}
\begin{tabular}{p{0.7cm}p{5.5cm}p{1.7cm}p{1.7cm}p{1.7cm}p{0.5cm}}
\toprule
\textbf{ID} & \textbf{Quote} & \textbf{1st Order Concepts} & \textbf{2nd Order Themes} & \textbf{Aggregate Dimensions} & \textbf{(\%)} \\
\midrule
R-66 & My colleagues don't influence on my decisions to use language models & No influence & Independent decision-making & No Influence & 29 \\ 
R-24 & Not at all. & No influence & Independent decision-making & No Influence & 29 \\ 
R-58 & Some are in favor, but it doesn't influence my opinion or way of using it & Others in favor, no influence on opinion or usage & Independent evaluation & No Influence & 29 \\ 
R-97 & Very much, as we use it on a daily basis and it is encouraged. & Use on a daily basis, encouraged & Strong adoption & High Influence & 26 \\ 
R-79 & We all use them, so we encourage each other to use them as well & Collective use, mutual encouragement & Collaboration and support & High Influence & 26 \\ 
R-27 & Greatly, everyone I know is keen on exploring their possibilities. & Strong influence, exploring possibilities & Enthusiastic adoption & High Influence & 26 \\ 
R-45 & Not much & Minimal influence & Low level of influence & Low Influence & 24 \\ 
R-99 & Not really. & Minimal influence & Low level of influence & Low Influence & 24 \\ 
R-9 & I make my own decisions ... but I often seek input and feedback from my colleagues and peers ... & Independent decision, input and feedback from peers & Valuing peer opinions & Moderate Influence & 21 \\ 
R-82 & There's some influence but I don't think it's a big deal because everyone uses it out of curiosity & Some influence, curiosity-driven adoption & Curiosity and exploration & Moderate Influence & 21 \\ 
\bottomrule
\end{tabular}
\label{tab:SCT-SI}
\end{table}

\subsection{Self-efficacy of LLMs in Software Engineering}

Self-efficacy, an integral construct within the Social Cognitive Theory, refers to an individual's belief in their ability to perform specific tasks and achieve desired outcomes~\cite{bandura1997SelfEfficacy}. In the context of software engineering, the self-efficacy of developers using Large Language Models can impact their adoption and effective utilization. Based on our thematic analysis, we identified three aggregate dimensions that contribute to understanding self-efficacy in relation to LLMs adoption: (1) Importance of being seen as cutting-edge, (2) Focus on practicality and efficiency, and (3) Low importance of being seen as cutting-edge. These dimensions will be discussed in detail in the following subsubsections, highlighting their role in shaping developers' self-efficacy and adoption behavior of LLMs in software engineering. The data structure can be found in Table \ref{tab:SCT-SE}.

\subsubsection{Importance of Being Seen as Cutting-edge}

As shown in the Table, 57\% of respondents emphasized the importance of being seen as someone who uses cutting-edge technology in their work. This perception aligns with the notion of self-efficacy, as developers who consider themselves proficient in the latest technologies tend to have higher confidence in their capabilities~\cite{compeau1995computer}. For example, R-2 expressed that being up-to-date with cutting-edge technology is crucial in their line of work, and R-30 mentioned the fear of being replaced by someone perceived as more proficient in cutting-edge technology. This focus on staying current with technological advancements may encourage developers to adopt LLMs to showcase their expertise and maintain a competitive edge in the industry~\cite{venkatesh2000theoretical}.\

\subsubsection{Focus on Practicality and Efficiency}

Our analysis revealed that 35\% of respondents highlighted the importance of practicality and efficiency in their work, demonstrating a preference for using technologies that effectively solve problems or meet client needs, rather than simply being cutting-edge. This focus reflects a more task-oriented self-efficacy, where developers concentrate on finding the most appropriate tools for the job. R-10, for instance, emphasized the importance of staying ahead of the curve and delivering innovative solutions to meet evolving customer needs, while R-19 and R-100 mentioned the balance between adopting cutting-edge technologies and ensuring efficiency in their work. This suggests that developers with a focus on practicality and efficiency will adopt LLMs only when they perceive these models to provide tangible benefits to their work.\

\subsubsection{Low Importance of Being Seen as Cutting-edge}

A smaller group of respondents (8\%) assigned low importance to being seen as using cutting-edge technology in their work. These developers prioritize stability, effectiveness, or other factors over adopting the latest technologies, which may influence their self-efficacy in terms of LLM adoption~\cite{torkzadeh2006contingency}. For instance, R-21 expressed a preference for stability over implementing bleeding-edge technology, and R-66 stated that their focus is on using the most effective tools for the task at hand, regardless of whether they are cutting-edge. This finding suggests that developers with a low emphasis on cutting-edge technology may be less inclined to adopt LLMs unless they demonstrate clear benefits over existing tools.
\\

In conclusion, our thematic analysis of self-efficacy in relation to LLM adoption in software engineering has identified three aggregate dimensions that provide valuable insights into developers' perceptions and behaviors. These dimensions suggest that the importance assigned to cutting-edge technology, along with the focus on practicality and efficiency, plays a critical role in shaping developers' self-efficacy and their likelihood of adopting LLMs.

\begin{table}[!ht]
\centering
\tiny
\caption{Data Structure of Self-efficacy of LLMs in Software Engineering}
\begin{tabular}{p{0.7cm}p{5.5cm}p{1.7cm}p{1.7cm}p{1.7cm}p{0.5cm}}
\toprule
\textbf{ID} & \textbf{Quote} & \textbf{1st Order Concepts} & \textbf{2nd Order Themes} & \textbf{Aggregate Dimensions} & \textbf{(\%)} \\
\midrule
R-2 & Being up-to-date with cutting-edge technology is crucial in my line of work. & Importance of cutting-edge & Emphasizing cutting-edge importance & Importance of being seen as cutting-edge & 57 \\

R-30 & Very important. I don't want to be replaced with someone who's seen like that while I don't. & Fear of being replaced & Emphasizing cutting-edge importance & Importance of being seen as cutting-edge & 57 \\

R-57 & It's important to show I'm capable of adapting. & Capability of adapting & Emphasizing cutting-edge importance & Importance of being seen as cutting-edge & 57 \\

R-77 & I think is really important to be always up to date with new technologies & Importance of being up to date & Emphasizing cutting-edge importance & Importance of being seen as cutting-edge & 57 \\

R-10 & it is important for me because it helps me to stay ahead of the curve and deliver innovative solutions that meet evolving customer needs. & Staying ahead, innovative solutions & Focus on practicality and efficiency & Focus on practicality and efficiency & 35 \\

R-19 & In my work as a software developer, it is important for me to keep up to date with the latest trends and technologies in software engineering, including language models. [...] & Meeting clients' needs, appropriate tech & Focus on practicality and efficiency & Focus on practicality and efficiency & 35 \\

R-47 & Not much, my work is mostly based on the technologies others decide to use. & Using decided technologies & Focus on practicality and efficiency & Focus on practicality and efficiency & 35 \\

R-100 & A little bit, but cutting-edge technology does not always mean higher efficiency. And efficiency is key. & Efficiency, not always cutting-edge & Focus on practicality and efficiency & Focus on practicality and efficiency & 35 \\

R-21 & Not at all. I'm looking for stability, not implementing the next bleeding-edge thing in my workflow. & Stability, rejecting cutting-edge tech & Low importance of being seen as cutting-edge & Low importance of being seen as cutting-edge & 8 \\

R-66 & I don't find it important to be seen as someone who uses cutting-edge technology in my work. My focus is on using the most effective tools for the task at hand. & Low importance of cutting-edge, effective tools & Low importance of being seen as cutting-edge & Low importance of being seen as cutting-edge & 8 \\
\bottomrule
\end{tabular}
\label{tab:SCT-SE}
\end{table}

\subsection{Environmental Factors of LLMs in Software Engineering}

The adoption of Large Language Models in software engineering is influenced by various environmental factors that shape organizational behaviors and decision-making processes. Drawing on the Social Cognitive Theory framework~\cite{bandura2014social}, this study investigates how environmental factors affect the extent to which organizations are supportive of adopting LLMs as a standard technology. The following sections present the findings of our thematic analysis, which revealed five aggregate dimensions related to environmental factors: Supportive Attitude, Neutral Stance, Conditional Support, Limited Support, and Lack of Support. Each dimension is discussed in detail with references to the data presented in Table \ref{tab:SCT-EF}.

\subsubsection{Supportive Attitude}

The most prevalent aggregate dimension in our data, Supportive Attitude, captures the positive and proactive stance of organizations in promoting LLM adoption (42\% frequency). This dimension encompasses strong organizational encouragement and investment in the technology to facilitate its integration into software engineering practices~\cite{rogers2010diffusion}. Respondent R-5, for instance, highlights the active promotion of LLMs by their organization, stating that they ``\textit{pay for it and encourage us to use it}.'' Similarly, R-45 reports a ``\textit{very supportive}'' attitude, illustrating the extent to which some organizations prioritize the adoption of LLMs.

\subsubsection{Neutral Stance}

The Neutral Stance dimension reflects organizations that neither actively support nor oppose the adoption of LLMs (20\% frequency). This stance may be attributed to the lack of awareness or knowledge about LLMs, or a wait-and-see approach to gauge the potential benefits and drawbacks of the technology~\cite{kohli2019digital}. Respondent R-84 describes their organization's position as ``\textit{neutral, up to the employee},'' implying that the decision to use LLMs is left to individual discretion rather than being guided by organizational policy.

\subsubsection{Conditional Support}

Some organizations in our sample exhibit a Conditional Support dimension (19\% frequency), characterized by their willingness to adopt LLMs provided certain criteria are met, such as the technology demonstrating clear benefits or aligning with specific organizational objectives~\cite{rogers2010diffusion}. In this context, R-26 notes that their organization supports LLM adoption ``\textit{if it brings more advantages to the team and the way we work, and make things faster}.''

\subsubsection{Limited Support}

The Limited Support dimension (12\% frequency) represents organizations that only support the use of LLMs in specific contexts or for certain tasks. This selective approach may stem from concerns related to security, privacy, or ethical considerations. For example, R-64 explains that their organization ``\textit{opposes them when working on new features but for debugging, they can be quite helpful}.''

\subsubsection{Lack of Support}

Finally, the Lack of Support dimension (15\% frequency) captures organizations that actively oppose or discourage the use of LLMs in software engineering. This stance may be driven by various factors, such as ethical concerns, fear of job displacement, or skepticism about the technology's effectiveness. Respondent R-74 reveals that their organization offers limited support for LLMs, mainly due to ``\textit{copyright and security concerns about proprietary intellectual property}.''
\\

In summary, our thematic analysis of environmental factors affecting LLM adoption in software engineering has identified five aggregate dimensions, ranging from strong support to active opposition. These dimensions provide valuable insights into the diverse organizational attitudes and contexts that shape the integration of LLMs into software engineering practices, contributing to a deeper understanding of the role of environmental factors in the SCT framework.

\begin{table}[!ht]
\centering
\tiny
\caption{Data Structure of Environmental Factors of LLMs in Software Engineering}
\begin{tabular}{p{0.7cm}p{5.5cm}p{1.7cm}p{1.7cm}p{1.7cm}p{0.5cm}}
\toprule
\textbf{ID} & \textbf{Quote} & \textbf{1st Order Concepts} & \textbf{2nd Order Themes} & \textbf{Aggregate Dimensions} & \textbf{(\%)} \\
\midrule

R-5 & A lot, they pay for it and encourage us to use it & encouragement, payment & active promotion & Supportive Attitude & 42 \\
R-45 & Very supportive & strong support & highly supportive & Supportive Attitude & 42 \\
R-84 & I think my organization is neutral, is up to the employee & neutral, employee choice & lack of strong opinion & Neutral Stance & 20 \\
R-26 & If it brings more advantages to the team and the way we work, and make things faster, they are supportive of adopting language models. & advantages, efficiency, conditional support & support with conditions & Conditional Support & 19 \\
R-21 & Not at all. We are anti-AI art, and also anti-AI when it comes to code. It failed to comply with our requirements anyway in testing. & not supportive, anti-AI & opposition to AI technology & Lack of Support & 15 \\
R-74 & Not much, since it might break copyright and also poses some security concerns about proprietary intellectual property & copyright concerns, security concerns, limited support & concerns leading to lack of support & Lack of Support & 15 \\
R-92 & I don't know, they don't oppose it though & uncertainty, no opposition & unsure & Uncertainty & 14 \\
R-61 & My organization is somewhat supportive but they are not sure & partial support, uncertainty & ambivalence & Uncertainty & 14 \\
R-79 & They're supportive of using it as a support tool, not as a main tool & support, limited context & limited application & Limited Support & 12 \\
R-64 & They oppose them when working on new features but for debugging, they can be quite helpful. & opposition, debugging, specific context & support in specific contexts & Limited Support & 12 \\
\bottomrule
\end{tabular}
\label{tab:SCT-EF}
\end{table}

\subsection{Key Insights of the Qualitative Study}
\label{sec:KeyInsights}

Our findings demonstrate the potential benefits of LLMs in software engineering, highlighting their impact on automating repetitive tasks, enhancing problem-solving abilities, facilitating learning and understanding, improving code quality, assisting in debugging and optimization, and increasing overall efficiency. However, concerns about the limitations of LLMs, such as their lack of knowledge about internal APIs and the need for human oversight, emphasize the importance of human expertise in utilizing these tools effectively.

The perceived ease of use of LLMs in software engineering is influenced by factors such as integration with existing tools and workflows, accessibility and comprehensibility of documentation, customizability and adaptability, and support from the developer community. These factors are crucial in facilitating the adoption of LLMs and their seamless integration into the software engineering domain.

The behavioral intention to adopt LLMs in software engineering is shaped by the perceived usefulness, perceived ease of use, social influence, and facilitating conditions. These factors collectively contribute to creating an environment conducive to the successful integration of LLMs into software engineering practices.

Regarding the Diffusion of Innovation theory, the compatibility of LLMs in software engineering is influenced by dimensions such as improved efficiency, assistance and support, similarity to current practices, and adaptation and learning. However, concerns about dependency and skill degradation, privacy, security and data protection, job displacement, and labor market implications, and accuracy, reliability, and explainability underline the complexity of LLM adoption in this domain.

Our findings also reveal the relative advantage of LLMs over traditional methods in software engineering. Factors such as time efficiency, code quality, user experience, learning and skill development, and customization and personalization contribute to the perceived benefits of LLMs in this domain.

From a Social Cognitive Theory perspective, the influence of peers and colleagues on LLM adoption in software engineering varies from no influence to high influence. Self-efficacy is influenced by factors such as the importance of being seen as cutting-edge, a focus on practicality and efficiency, and the low importance of being seen as cutting-edge. Environmental factors, such as supportive organizational culture, uncertainty, security concerns, neutral organizational culture, resistance to change, and limited or marginal support, also play a role in LLM adoption.

In conclusion, the adoption of LLMs in software engineering is a multifaceted phenomenon influenced by a variety of factors and theoretical perspectives. Our study provides valuable insights into the key dimensions and concerns that shape the integration of LLMs in the software engineering domain, paving the way for future research and practice in this area.

\section{The Human-AI Collaboration and Adaptation Framework (HACAF)}
\label{sec:HACAF}

In this section, we introduce the Human-AI Collaboration and Adaptation Framework (HACAF), an innovative theoretical model designed to understand and predict the adoption of Generative AI tools in software engineering. The HACAF derives its components from several established theories including the Technology Acceptance Model, Diffusion of Innovation Theory, Social Cognitive Theory, the Unified Theory of Acceptance and Use of Technology (UTAUT), and the personal innovativeness construct.

While the original TAM, DOI, and SCT theories provide robust theoretical foundations for understanding technology acceptance and adoption, our qualitative investigation indicated the necessity for a more nuanced model. The HACAF is, therefore, not merely an amalgamation of these theories, but also an evolution, as it incorporates additional facets revealed in our research.

The inclusion of constructs from UTAUT addresses the need for a greater focus on social influence and facilitating conditions, elements that emerged as significantly influential in our qualitative findings. The additional integration of the personal innovativeness construct into HACAF is motivated by the observed variability in adoption behaviors among software engineers, even within the same contextual environment, implying a role for individual differences in innovative tendencies.

By merging these four main components into the HACAF, we not only leverage the collective strengths of these prominent theories but also account for the additional complexities of technology adoption that surfaced in our qualitative investigation. Consequently, the HACAF represents a tailored approach that reflects the multifaceted nature of LLM adoption in software engineering. This comprehensive framework aims to provide a deeper understanding of the complex dynamics involved in the adoption of LLMs and act as a guide for future research and practice in this rapidly evolving domain.

\textbf{Perceptions about the technology} is a cornerstone of HACAF, rooted in TAM. This construct denotes a software engineer's evaluation of the usefulness, ease of use, and relative advantage of LLMs. The qualitative data substantiates this construct by revealing that software engineers assess LLMs based on their potential to streamline coding processes, enhance code quality, and expedite project timelines. Furthermore, the focus on practicality and efficiency found in our study underscores the importance of this construct in influencing the adoption of LLMs.

\textbf{Compatibility factors}, informed by the Diffusion of Innovation theory, illustrate the degree to which LLMs align with the existing values, experiences, and needs of potential adopters. Our qualitative investigation underpins this construct by highlighting the importance of the fit of LLMs within current software development workflows. Developers who perceive a high degree of fit, irrespective of the technology's cutting-edge nature, are more likely to adopt LLMs, reinforcing the relevance of this construct.

\textbf{Social factors}, drawing on UTAUT and the concept of computer self-efficacy, emphasize the influence of the social environment and an individual's belief in their abilities to use LLMs. The investigation's findings lend weight to this construct by indicating that the importance of being seen as cutting-edge and the developer's self-efficacy could drive LLM adoption. The role of peer approval and the belief in one's competence to master LLMs surfaced as significant influencers, further establishing this construct's salience in the HACAF model.

\textbf{Personal and environmental factors} bring into play the role of personal innovativeness and organizational support. Personal innovativeness represents an individual's predisposition towards new technologies, and organizational support captures the perceived facilitating conditions within an organization for the use of LLMs. Both elements emerged as significant in our investigation. Our data evidenced how individual readiness to experiment with LLMs and the organization's stance towards LLMs, ranging from active support to outright opposition, directly influence the likelihood of LLM adoption.

In conclusion, the HACAF model, informed and justified by our qualitative investigation and underpinned by established theoretical constructs, offers a nuanced understanding of the interplay of individual and organizational factors influencing the adoption of LLMs in software engineering. By grounding this framework in both empirical evidence and theory, we aim to provide a solid foundation for further empirical scrutiny and contribute to the understanding of AI technology adoption dynamics.

\subsection{Theoretical Model and Hypotheses}

\begin{figure}[ht!]
\centerline{\includegraphics[width=1\linewidth]{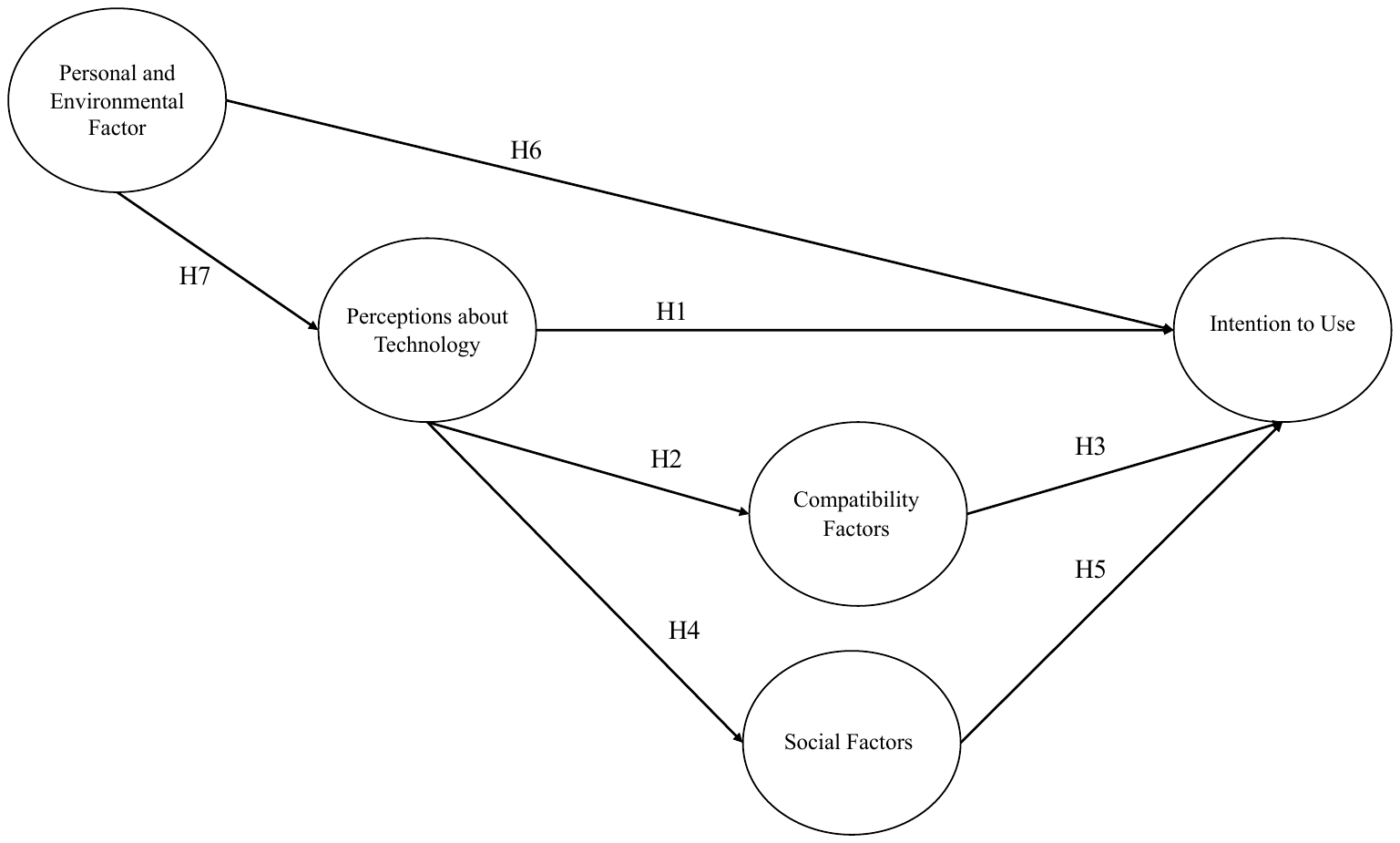}}
\caption{The Human-AI Collaboration and Adaptation Framework (HACAF)}
\label{fig:HACAF}
\end{figure}

The previous section's theoretical foundation, bolstered by the qualitative investigation, serves as the basis for operationalizing the Human-AI Collaboration and Adaptation Framework (HACAF). We now translate these theoretical constructs into four main determinants of the intention to use LLMs: perceptions about the technology, compatibility factors, social factors, and personal and environmental factors, and formulate hypotheses based on these constructs, graphically represented in Figure \ref{fig:HACAF}.

\textbf{Perceptions about the technology} encapsulate the perceived usefulness, ease of use, and relative advantage of LLMs. As our qualitative data revealed, LLMs' perceived usefulness and relative advantage, such as streamlining coding processes and enhancing code quality, have a significant bearing on their adoption. The ease of use was another critical factor, as intuitive and user-friendly LLMs are more likely to be adopted by software engineers. Thus, drawing upon the Technology Acceptance Model, we propose:

\begin{hypothesis}
H1: Positive perceptions about the technology (PT), encompassing perceived usefulness, ease of use, and relative advantage, will increase the Intention to Use (IU) LLMs in a software engineering context.
\end{hypothesis}

\textbf{Compatibility factors} address the extent to which LLMs align with the existing values, experiences, and needs of potential adopters. As the qualitative investigation underlines, LLMs' compatibility with current software development workflows significantly influences adoption decisions. Software engineers are more likely to adopt LLMs when they perceive a high degree of fit with their work practices. Following this and the Diffusion of Innovation theory, we hypothesize:

\begin{hypothesis}
H2: Positive perceptions about the technology (PT) will enhance the Compatibility Factors (CF).
H3: Enhanced compatibility factors (CF) will in turn increase the Intention to Use (IU) LLMs.
\end{hypothesis}

\textbf{Social factors}, which include social influence and self-efficacy, play an important role in adoption decisions. Our qualitative study emphasized that peer approval and an individual's belief in their ability to use LLMs could significantly influence the intention to use these technologies. Based on this and the Unified Theory of Acceptance and Use of Technology (UTAUT) and the concept of computer self-efficacy, we posit:

\begin{hypothesis}
H4: Positive perceptions about the technology (PT) will increase Social Factors (SF).
H5: Increased social factors (SF) will enhance the Intention to Use (IU) LLMs.
\end{hypothesis}

Finally, \textbf{Personal and environmental factors}, including personal innovativeness and organizational support, contribute to the complexity of the model. As underscored by our investigation, an individual's willingness to experiment with LLMs and the perceived supportiveness of the organization can be decisive for LLM adoption. Thus, we hypothesize:

\begin{hypothesis}
H6: Personal and Environmental Factors (PEF), specifically personal innovativeness and organizational support, will moderate the relationship between Perceptions about the technology and IU LLMs, strengthening the positive effect of Perceptions about the technology on Intention to Use LLMs.
\end{hypothesis}

\section{Theory validation}
\label{sec:QuantStudy}

In the subsequent phase of our research, we transitioned from the qualitative insights garnered in the initial study to a quantitative validation. The qualitative findings from our initial study played a foundational role in shaping the subsequent phases of our research. 
Also here, we used the SIGSOFT Empirical Standard for Questionnaire Surveys and Multi-Methodology and Mixed Methods Research~\cite{ralph2020empirical}.

\textbf{Partial Least Squares -- Structural Equation Modeling (PLS-SEM):} This method was chosen based on its ability to validate complex models, especially when the research is in an exploratory stage, as is the case with our study on the adoption dynamics of Generative AI tools in software engineering.

\textbf{Scale Development:} The themes and patterns identified in our qualitative findings were instrumental in developing the scales for our quantitative study. These scales were designed not only to encapsulate the core of our qualitative insights but also to render them into measurable metrics. In line with established methodological guidelines~\cite{russo2021pls}, we adapted existing scales to ensure robustness and relevance. A detailed breakdown of these scales can be found in Table~\ref{tab:Items}.

\textbf{Survey Data Collection:} The design of our survey was directly informed by the conclusions drawn from our qualitative exploration. Questions were framed to test the hypotheses that emerged from the qualitative study, ensuring a coherent flow between the two research phases.

\textbf{Participant Demographics:} The demographics were chosen based on the insights from the qualitative study to ensure that the quantitative study sample was representative of the broader population of software engineers who might be impacted by the adoption of Generative AI tools.

\textbf{Evaluation of the Measurement Model:} The constructs used in the measurement model were derived from the themes of the qualitative study. This ensured that the quantitative analysis was grounded in the real-world experiences and perceptions of our initial study participants.

\textbf{Evaluation of the Structural Model:} The relationships tested in the structural model were informed by the patterns and relationships identified in the qualitative study. This ensured that our quantitative validation was directly aligned with the insights from our initial exploration.

\subsection{Partial Least Squares -- Structural Equation Modeling}
\label{ssec:PLS}

Partial Least Squares -- Structural Equation Modeling (PLS-SEM) is a multifaceted statistical examination designed to substantiate latent and unseen variables (or constructs) through multiple observable indicators. This method is particularly useful for theory development studies and is increasingly adopted in empirical software engineering~\cite{russo2021pls}. PLS-SEM can address multiple interconnected research queries in one extensive analysis, making it a popular choice across other fields such as Management, Information Systems Research, and Organizational Behavior. As Gefen et al. suggest, SEM is commonly employed to authenticate tools and verify connections between constructs~\cite{gefen2000structural}. The subsequent evaluation and analysis of the PLS-SEM model adhere to the latest guidelines and recommendations for research in software engineering by Russo \& Stol~\cite{russo2021pls}.

\subsubsection{Scale Development}

The survey was crafted with the assistance of supplementary theory. We structured our survey by adapting instruments from previous research. All items utilized to define each construct and the references used to shape the questions are summarized in Table \ref{tab:Items}. Each construct was assessed through uni-dimensional items on a 7-point Likert scale indicating levels of agreement.

Initially, a pre-test was conducted with three potential respondents (software engineers) to assess the survey's usability, reasoning, and phrasing. The usability received positive feedback, and some minor issues with the reasoning and phrasing were identified and subsequently addressed.

\subsubsection{Survey Data Collection}

The minimum sample size was determined by conducting an \textit{a priori} power analysis with G*Power. With an effect size of 15\%, significance level at 5\%, and a power of 95\%, the smallest size required for seven predictors is 153.

A cluster sampling strategy was utilized for data collection, facilitated through the academic data collection platform, Prolific. The survey was delivered via Qualtrics, randomizing the order of questions within their blocks to reduce response bias.

A multi-phase screening process was implemented to ensure the integrity of the collected data. Data collection was carried out between April and June 2023.

\textbf{Pre-screening:} During the initial selection phase, participants were chosen based on specific self-reported criteria. These included proficiency in \textit{Computer Programming}, employment in the \textit{Software} industry on a \textit{Full-time} basis, a non-student status, and an \textit{Approval rate} of 100\%. On Prolific, the approval rate represents the percentage of a participant's submissions that have been approved by researchers, indicating their reliability and consistency in providing quality responses. To emphasize our target demographic, our survey was titled ``[SOFTWARE ENGINEERS ONLY] Human-AI Collaboration and Adaptation.'' This title was chosen to clearly communicate to professionals who met the pre-screening criteria but were not actively working as software engineers. In total, 831 potential participants met these criteria.

\textbf{Competence Screening.} To ensure the accuracy of our pool, participants were asked to complete a questionnaire containing three competency-based questions about software design and programming. 
Additionally, they conveyed their familiarity with, and usage of, Generative AI tools, at least to a certain extent.
This helped confirm the reliability of their self-reported skills. This process resulted in a reduced pool of 606 participants.

\textbf{Quality \& Competence Screening.} 
To ensure the accuracy of our pool, we added one single-item screening question in our survey from Danilova et al.~\cite{danilova2021you}, asking ``\textit{Which of these websites do you most frequently use as aid when programming?}'' with `Stack Overflow' as the correct question. 
Additionally, we added three random attention checks to further ensure data quality.
A total of 220 completed questionnaires were received, but 36 were discarded due to failure on at least one attention check. This left us with 184 valid and complete responses, surpassing the minimum sample size.

\subsubsection{Participant Demographics}


Our survey encompassed a diverse set of 184 respondents, comprising 80\% males, 18\% females, 1\% non-binary individuals, and 1\% who preferred not to disclose their gender. The respondents were drawn from a broad geographic pool spanning 27 unique countries, with the most populous responses originating from the UK (24\%), South Africa (13\%), Poland (11\%), Germany (11\%), and the United States of America (7\%).

In terms of work tenure, the median experience among participants was three years. The majority, 125 respondents, were relatively early in their careers, with 1 to 5 years of experience. Forty participants reported a more substantial work experience ranging between 6 to 15 years. An additional 14 participants had an extensive work experience of 16 to 30 years, and a handful of respondents, 5 in total, possessed more than 30 years of experience.

Our sample prominently featured individuals from the software development sector, making up 66\% of all respondents. Additionally, 12\% of respondents held data analysis, engineering, or science roles. A smaller segment, 8\%, held leadership roles such as Team Leads or CIOs. The remaining respondents included Tester / QA Engineers (6\%), DevOps/Infrastructure Engineers (3\%), Architects (2\%), UX/UI Designers (2\%), and other roles (2\%).

\subsection{Evaluation of the Measurement Model}

In order to ensure the validity and reliability of our structural model, it is paramount to evaluate the reliability of the latent variables. Consequently, we analyze into the discriminant validity, internal consistency reliability, and convergent validity initially. 

\subsubsection{Discriminant Validity}

In this context, discriminant validity refers to the distinctness or uniqueness of one latent variable compared to another. This serves as an essential parameter for determining whether two constructs are essentially the same and represent varying facets of knowledge. For its evaluation, we utilized the Heterotrait-Monotrait ratio of correlations (HTMT) which is recognized for its superior performance over other tests like the Fornell-Larcker criterion. The HTMT values should ideally be below 0.90.

\begin{table}[h]
\centering
\small
\robustify{\textbf}
\caption{Heterotrait-Monotrait ratio of correlations (HTMT) of the model}
\label{tab:HTMT}
    \begin{tabular}{p{6cm} p{1cm} p{1cm} p{1cm} p{1cm} p{1cm}}
        \toprule
  & CF & IU & PEF & PT \\
\hline
Compatibility Factors (CF) &  &  &  & \\
Intention to Use (IU) & 0.756 &  &  & \\
Personal and Environmental Factors (PEF) & 0.372 & 0.272 &  & \\
Perceptions about the Technology (PT) & 0.848 & 0.633 & 0.338 & \\
Social Factors (SF) & 0.403 & 0.371 & 0.441 & 0.437  \\
    \bottomrule
\end{tabular}
\end{table}

Table \ref{tab:HTMT} reveals that all coefficients fall beneath the predefined threshold, which suggests that every construct in the model represents a distinct phenomenon.

\subsubsection{Internal Consistency Reliability}

This test seeks to confirm that the items are gauging the latent variables in a consistent and reliable manner. As such, we refer to the Cronbach's Alpha, rho\_a, and rho\_c values showcased in Table \ref{tab:ICR}, all of which should exceed 0.60~\cite{nunnally1978psychometric}. 
We can conclude that our tests meet the reliability criterea.

\begin{table}[h]
\centering
\small
\robustify{\textbf}
\caption{Internal consistency reliability}
\label{tab:ICR}
    \begin{tabular}{p{6cm} p{1.5cm} p{1cm} p{1cm} p{1cm} }
        \toprule
 & Cronbach's alpha & rho\_a & rho\_c & AVE \\
\midrule
Compatibility Factors (CF) & 0.856 & 0.866 & 0.902 & 0.699 \\
Intention to Use (IU) & 0.939 & 0.941 & 0.961 & 0.891 \\
Personal and Environmental Factors (PEF) & 0.876 & 0.905 & 0.914 & 0.727 \\
Perceptions about the Technology (PT) & 0.948 & 0.950 & 0.956 & 0.685 \\
Social Factors (SF) & 0.875 & 0.906 & 0.940 & 0.887 \\
\bottomrule
\end{tabular}
\end{table}

\subsubsection{Convergent Validity}

The final validity assessment examines the extent of correlations between various items and their corresponding construct. It is noteworthy that our latent variables are reflectively measured (Mode A)\footnote{For a comprehensive comparison between reflective and formative measures, see Russo \& Stol (2021).}. As a result, these indicators should demonstrate a substantial variance proportion by converging on their latent variables. Two tests were employed to verify this assumption.

The first test involves the average variance extracted (AVE), which should register a value exceeding 0.5~\cite{hair2016pls}. The second test involves ensuring that the outer loadings of each measurement model for the latent variable account for at least 50\% variance. This is tested by assessing the indicator's reliability, which should exceed the square root of 50\%, i.e., 0.7.

Table \ref{tab:OuterLoadings} encapsulates the results of the indicator's reliability using cross-loadings. The items that did not contribute significantly to the variance and were subsequently excluded from our model during the analyis phase (a complete list of excluded items can be found in Table \ref{tab:Items}. Consequently, an improvement in the AVE was noted, thereby reinforcing the model's robustness.

\begin{table}[h]
\centering
\small
\robustify{\textbf}
\caption{Cross loadings (full list of items in Table \ref{tab:Items})}
\label{tab:OuterLoadings}
\begin{tabular}{p{1cm} p{2cm} p{2cm} p{2cm} p{2cm} p{2cm} }
        \toprule
Item & CF & IU & PEF & PT & SF \\
\midrule
CF2 & \textbf{0.896} & 0.626 & 0.293 & 0.686 & 0.333 \\
CF3 & \textbf{0.856} & 0.657 & 0.352 & 0.666 & 0.305 \\
CF5 & \textbf{0.775} & 0.482 & 0.248 & 0.547 & 0.250 \\
CF6 & \textbf{0.811} & 0.505 & 0.236 & 0.654 & 0.289 \\
IU1 & 0.662 & \textbf{0.935} & 0.238 & 0.560 & 0.293 \\
IU2 & 0.602 & \textbf{0.945} & 0.235 & 0.558 & 0.337 \\
IU3 & 0.671 & \textbf{0.951} & 0.267 & 0.577 & 0.330 \\
PEF4 & 0.293 & 0.198 & \textbf{0.811} & 0.276 & 0.277 \\
PEF5 & 0.156 & 0.130 & \textbf{0.838} & 0.214 & 0.250 \\
PEF6 & 0.362 & 0.316 & \textbf{0.892} & 0.300 & 0.394 \\
PEF7 & 0.298 & 0.200 & \textbf{0.867} & 0.258 & 0.391 \\
PT1 & 0.658 & 0.581 & 0.251 & \textbf{0.841} & 0.340 \\
PT11 & 0.641 & 0.521 & 0.178 & \textbf{0.855} & 0.318 \\
PT12 & 0.490 & 0.415 & 0.362 & \textbf{0.703} & 0.350 \\
PT13 & 0.639 & 0.491 & 0.295 & \textbf{0.856} & 0.327 \\
PT14 & 0.612 & 0.460 & 0.267 & \textbf{0.851} & 0.330 \\
PT2 & 0.663 & 0.545 & 0.281 & \textbf{0.827} & 0.378 \\
PT3 & 0.676 & 0.503 & 0.262 & \textbf{0.854} & 0.344 \\
PT4 & 0.655 & 0.511 & 0.271 & \textbf{0.874} & 0.383 \\
PT6 & 0.669 & 0.550 & 0.234 & \textbf{0.866} & 0.323 \\
PT7 & 0.622 & 0.353 & 0.191 & \textbf{0.727} & 0.243 \\
SF1 & 0.295 & 0.291 & 0.388 & 0.325 & \textbf{0.929} \\
SF2 & 0.365 & 0.342 & 0.359 & 0.427 & \textbf{0.955} \\
\bottomrule
\end{tabular}
\end{table}

\subsection{Evaluation of the Structural Model}

\begin{figure}[ht!]
\centerline{\includegraphics[width=1\linewidth]{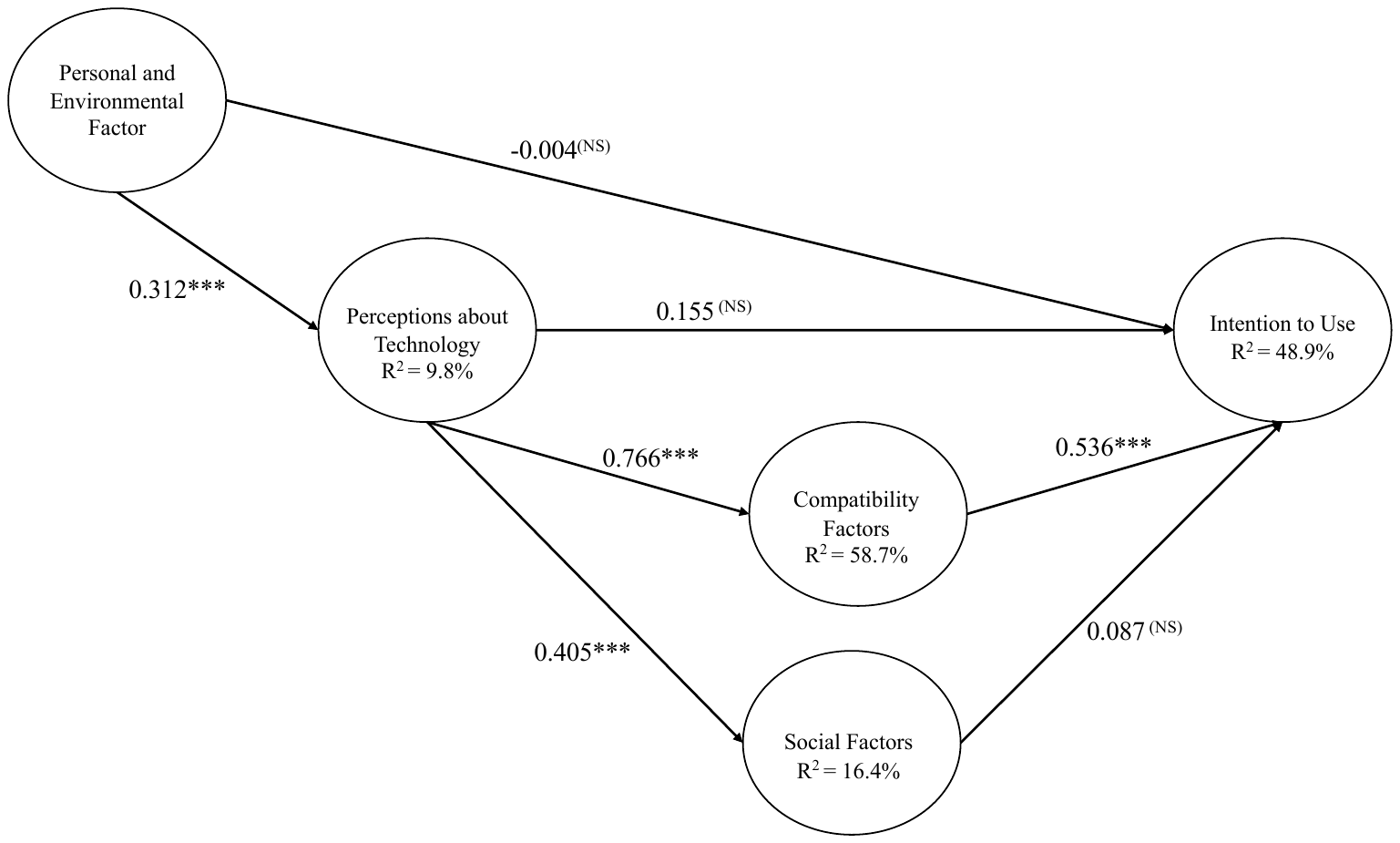}}
\caption{HACAF's structural model with $R^2$ and path coefficients (*** \textit{p}<0.001, (NS) \textit{p}>0.05).}
\label{fig:HACAF-SM}
\end{figure}

After ensuring the reliability of all our constructs through our Measurement Model assessment, we can now shift our focus towards evaluating the Structural Model, graphically represented in Figure~\ref{fig:HACAF-SM}. This evaluation is pivotal in discussing the predictive power of our model and validating our research hypotheses.

\subsubsection{Collinearity Analysis}

Initially, we analyze the correlation between the exogenous variable (Personal and environmental factors) and other endogenous variables. These should be independent to prevent any potential bias in the path estimations. The Variance Inflation Factor (VIF) test, which detects multicollinearity (i.e., an extreme degree of collinearity), should have a value under five~\cite{Miles2014VIF}. Our VIF values are below this threshold, ranging from 4.710 (for the IU\_3 item) to 2.034 (PEF\_4). Consequently, we deduce that our model doesn't suffer from multicollinearity issues.

\subsubsection{Path Relations: Significance and Relevance}

Path coefficients represent the hypothesized relationships between latent variables. They are standardized, meaning their values can range from -1 to +1. PLS-SEM does not require distributional assumptions, which means we can not use parametric tests to assess significance. As a workaround, we use a two-tailed bootstrapping method that incorporates 5,000 subsamples with replacement. 

Details of the bootstrapping results are presented in Table \ref{tab:PathCoeff}, which includes the bootstrapping coefficients, mean, standard deviation, T statistics, and p-values corresponding to each of our seven hypotheses.

Our analysis reveals that a majority of our hypotheses are statistically significant. Specifically, four relationships have p-values less than 0.05, and their T statistics surpass 1.96, indicative of a 5\% significance level as noted by Hair et al.

\begin{table}[h]
\centering
\small
\robustify{\textbf}
\caption{Path coefficients, bootstrap estimates, standard deviation, T statistics, and \textit{p}-values}
\label{tab:PathCoeff}
\begin{tabular}{p{4cm} p{2cm} p{2cm} p{1cm} p{1cm} p{1cm} }
        \toprule
Hypothesis & Coefficient  & Bootstrap Mean & St.Dev. & T & \textit{p}  \\
\midrule
\textit{H1}: PT $\rightarrow$ IU & 0.155 & 0.150 & 0.122 & 1.271 & 0.204\\
\textit{H2}: PT $\rightarrow$ CF & 0.766 & 0.766 & 0.047 & 16.405 & 0.000\\
\textit{H3}: CF $\rightarrow$ IU & 0.536 & 0.537 & 0.090 & 5.936 & 0.000\\
\textit{H4}: PT $\rightarrow$ SF & 0.405 & 0.408 & 0.076 & 5.346 & 0.000\\
\textit{H5}: SF $\rightarrow$ IU & 0.087 & 0.090 & 0.064 & 1.373 & 0.170\\
\textit{H6}: PEF $\rightarrow$ IU & -0.004 & -0.004 & 0.056 & 0.064 & 0.949\\
\textit{H7}: PEF $\rightarrow$ PT & 0.313 & 0.317 & 0.072 & 4.313 & 0.000\\
\bottomrule
\end{tabular}
\end{table}

\subsubsection{Evaluation of Determination Coefficients}

After affirmatively determining the significance of the majority of our hypotheses, we now proceed to the final phase of our study, which centers on the predictive power of the endogenous constructs. This is shown in Table \ref{tab:R2}. The predictive capacity is quantified through the variance explained ($R^2$) by the endogenous constructs. $R^2$ signifies the ratio of the variance in the dependent variable that can be predicted from the independent variables. As $R^2$ increases with the number of predictors (more predictors equate to a higher $R^2$), it is prudent to consider the adjusted $R^2$ as well, which takes into account the quantity of predictors in the model. The values for these metrics lie between 0 and 1. Determining a standard for $R^2$ can be challenging as its significance heavily relies on the topic at hand~\cite{hair2016pls}, but it is generally accepted that it should exceed 0.19~\cite{chin1998PLS}.

\begin{table}[h]
\centering
\small
\robustify{\textbf}
\caption{Coefficients of determination}
\label{tab:R2}
\begin{tabular}{p{6cm} p{2cm} p{2cm} }
        \toprule
Construct & $R^2$        & $R^2$ Adjusted    \\
\midrule
Compatibility Factors  & 0.587 & 0.585 \\
Intention to Use  & 0.489 & 0.477 \\
Personal and Environmental Factors & 0.098 & 0.093 \\
Social Factors  & 0.164 & 0.159 \\
\bottomrule
\end{tabular}
\end{table}

\subsubsection{Evaluating Predictive Efficacy}

Given that our study's primary objective is to ascertain predictions rather than causality, we undertook a predictive results evaluation employing PLSpredict \cite{shmueli2016elephant}. This approach tests if a model (developed using a training sample) can predict the outcomes from a test sample. Our sample was segmented into ten parts, and ten repetitions were utilized to derive the PLSpredict statistics. The results were interpreted based on the guidelines proposed by Shmueli et al. \cite{shmueli2019predictive}. Table \ref{tab:PLSpredict} portrays the latent variables' predictive accuracy. Notably, all variables display a strongly positive \textit{Q}\textsuperscript{2}$_{\text {predict}}$ indicating robust performance of the model.

\begin{table}[h]
\centering
\small
\robustify{\textbf}
\caption{Constructs prediction summary}
\label{tab:PLSpredict}
\begin{tabular}{p{6cm} p{2cm} p{2cm} p{2cm}}
        \toprule
Construct & RMSE            & MAE   & \textit{Q}\textsuperscript{2}$_{\text {predict}}$       \\
\midrule
Compatibility Factors  & 0.089 & 0.984 & 0.752 \\
Intention to Use  & 0.046 & 1.012 & 0.761 \\
Perceptions about the Technology  & 0.076 & 0.998 & 0.733 \\
Social Factors  & 0.078 & 0.971 & 0.755 \\
\bottomrule
\end{tabular}
\end{table}

\subsubsection{Assessing Predictive Consistency}

Our final examination focuses on the predictive consistency of our model, for which we scrutinize the effect sizes ($f^2$) as illustrated in Table \ref{tab:f2}. This assessment involves understanding the impacts of various relationships within the model. The threshold values for effect sizes are set at 0.02, 0.15, and 0.35, corresponding to small, medium, and large effects, respectively~\cite{Cohen1988PowerAnalysis}. Here, the relationship with the strongest effect are compatibility factors with intention to use LLMs\footnote{Although larger effect sizes are not inherently problematic, they can occasionally suggest a potential risk of overfitting. However, we have performed a comprehensive examination of this potential issue in Appendix \ref{sec:Overfitting} and concluded that overfitting is not present in our model.}.

\begin{table}[h]
\centering
\small
\robustify{\textbf}
\caption{Effect sizes ($f^2$)}
\label{tab:f2}
\begin{tabular}{p{5cm} p{1cm} p{1cm} p{1cm} p{1cm} p{1cm}}
        \toprule
Constructs & CF & IU & PEF & PT & SF \\
\midrule
Compatibility Factors   &     & 0.225 &     &     &     \\
Intention to Use   &     &     &     &     &     \\
Personal and Environmental Factors  & 0.000 &     & 0.108 &     &     \\
Perceptions about the Technology   & 1.426 &     &     &     & 0.196 \\
Social Factors   & 0.018 & 0.011 &     &     &     \\
\bottomrule
\end{tabular}
\end{table}

\subsection{Determining Key Factors for Adoption: An Analysis using Importance-Performance Map}

This focused study specifically explores the elements influencing the adoption and intended use of Generative AI tools in the field of software engineering. Using the Importance-Performance Map Analysis (IPMA) methodology, we have combined the analysis of both the importance and performance dimensions derived from the PLS-SEM investigation \cite{Ringle2016IPMA}. This approach allows us to determine the extent to which various constructs contribute to the enhancement of the target construct - in this case, the Intention to Use (IU) and Adoption of Generative AI tools. It provides strategic insights by identifying which constructs are most significant and which ones demand improvements in performance.

Table \ref{tab:LV_Performance} shows that all identified constructs, namely Compatibility Factors (CF), Personal and Environmental Factors (PEF), Perceptions about the Technology (PT), and Social Factors (SF), demonstrate robust performance, all exceeding 65\%, with PT and CF even surpassing the 83\% mark. This result is noteworthy, especially considering that established models such as the technology acceptance model typically show a constructs' performance range between 50\% and 70\% \cite{ramkumar2019q}.

The importance of individual constructs as shown in Table \ref{tab:UnTotalEffects} is consistent with those of mature models, with values between 0.110 and 0.767. In particular, PT and CF emerge as the most significant constructs influencing IU. This finding emphasizes the role of perceptions about the technology and compatibility factors in the intention to use and the eventual adoption of Generative AI tools in software engineering.

\begin{figure}[!ht]
\centering
\includegraphics[width=0.8\linewidth]{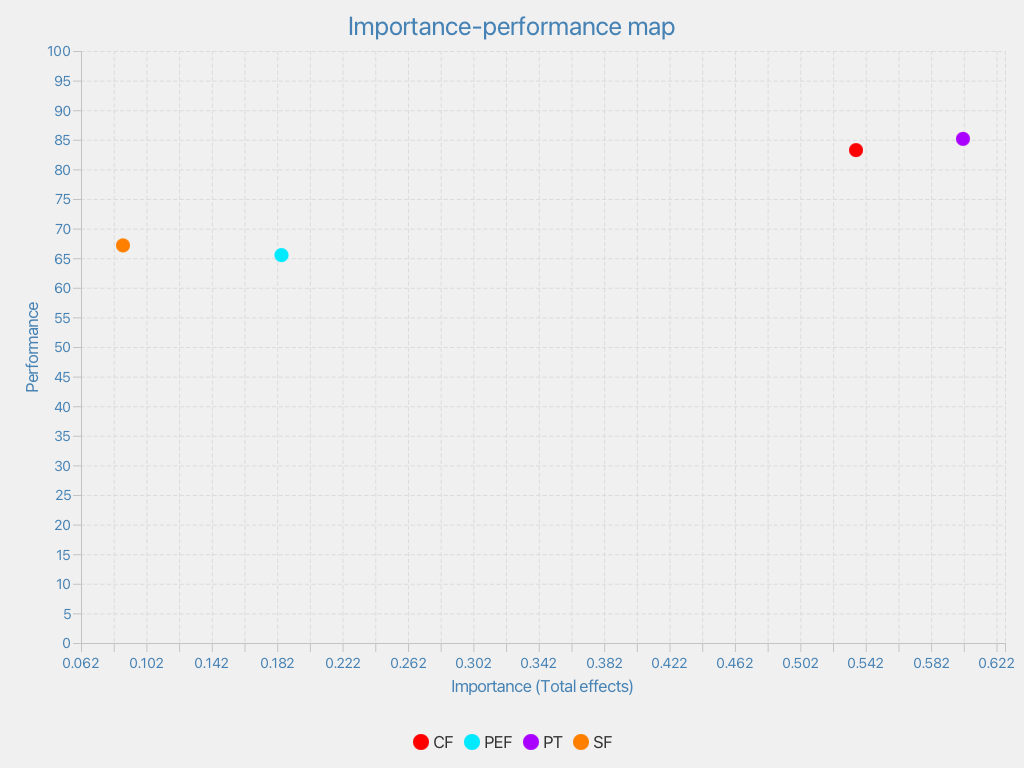}
\caption{Importance-Performance Map Analysis of Intention to Use.}
\label{fig:IPMA}
\end{figure}

Figure \ref{fig:IPMA} provides a visualization of the interplay between the importance and performance of these constructs. For instance, a unit increase in the performance of PT (from, say, 85.138 to 86.138) would improve the IU by the total effect of PT on IU, which is 0.540. This suggests that if the goal is to increase IU and Adoption, emphasis should be placed on enhancing PT, given its high importance. Similarly, also CF play a crucial role to support AI adoption. On the other hand, constructs such as SF and PEF, despite their role, appear less critical to the intention to use and adoption of Generative AI tools.

\begin{table}[!ht]
\centering
\sisetup{
    group-digits=true,
    group-minimum-digits=4,
    table-format=0.3,
    mode=text,
    detect-weight=true, 
    detect-family=true
}
\small
\robustify{\textbf}
\caption{Constructs Performance referred to Project Success}
\label{tab:LV_Performance}
\begin{tabular}{p{4cm} p{4cm}}
        \toprule
Construct & Construct Performances        \\
    \midrule
CF & 83.257 \\ 
PEF & 65.519 \\
PT & 85.138 \\
SF & 67.161 \\
\bottomrule
\end{tabular}
\end{table}

\begin{table}[!ht]
\centering
\sisetup{
    group-digits=true,
    group-minimum-digits=4,
    table-format=0.3,
    mode=text,
    detect-weight=true, 
    detect-family=true
}
\small
\robustify{\textbf}
\caption{Constructs Importance (Unstandardized Total Effects)}
\label{tab:UnTotalEffects}
\begin{tabular}{p{2cm} p{1cm} p{1cm} p{1cm} p{1cm} p{1cm}}
        \toprule
Construct & CF & IU & PEF & PT & SF \\
    \midrule
CF & & 0.646 & & & \\ 
IU & & & & & \\
PEF & 0.240 & 0.167 & & 0.313 & 0.127 \\
PT & 0.767 & 0.540 & & & 0.405 \\
SF & & 0.110 & & & \\
\bottomrule
\end{tabular}
\end{table}

\section{Discussion}
\label{sec:Discussion}

In our investigation of Generative AI adoption in software engineering, we developed the Human-AI Collaboration and Adaptation Framework (HACAF) to dissect the interplay between perceptions about the technology, compatibility factors, social factors, and personal and environmental factors. We did not to restrict our study to a specific generative AI model, such as ChatGPT-4 or ChatGPT3.5. This decision was driven by our aim to capture a holistic understanding of the adoption dynamics, recognizing that the landscape of generative AI is rapidly evolving. Committing to a single model could have limited the generalizability and longevity of our findings. 
By adopting a broader approach, we hypothesize that our framework may offer valuable insights, potentially remaining relevant as new iterations and variations of generative AI models emerge in the field, though this supposition invites further empirical validation. The results revealed a complex landscape, with surprising deviations from traditional technology acceptance theories.
Table~\ref{tab:Implications} summarizes our key findings and implications.

\begin{table}[!ht]
\centering
\Small
\robustify{\textbf}
\caption{Summary of findings and implications}
\label{tab:Implications}
\begin{tabularx}{\textwidth}{XXX}
        \toprule
Hypothesis & Findings  &  Implications   \\
    \midrule
\textit{H1}: Perceptions about the technology $\rightarrow$ Intention to Use & Not supported. We did not find a direct relationship between Perceptions about the technology and Intention to Use. &  Perception of the technology alone does not instigate the adoption of a Generative AI tool.  \\
   \addlinespace
\textit{H2}: Perceptions about the technology $\rightarrow$ Compatibility Factors & Supported. This relationship is the strongest of the model with a path coefficient of 0.77 with a very large effect size (1.43). & Assimilating the capabilities of Generative AI tools and their potential integration into existing software development workflows is crucial. \\
   \addlinespace
\textit{H3}: Compatibility Factors $\rightarrow$ Intention to Use & Supported. This relation is symmetrical to the H2 hyphotesis (with a considerable path coefficient of 0.54 and a large effect size  of 0.66), which mostly explains the intention to use of LLMs. The IPMA shows that Compatibility Factors are the most important element to improve the Intention to Use. Additionally, they explain, almost single handed, the adoption of AI tools with a very high $R^2$ close to 50\%. & The adoption of Generative AI tools hinges largely on their successful integration with existing software development workflows. \\
   \addlinespace
\textit{H4}: Perceptions about the technology $\rightarrow$ Social factors & Supported. We report a substantial path coefficient (0.40) with a medium effect size (0.20). The way Generative AI is perceived positively influence developer’s self-efficacy. & Developers consider LLMs as vital facilitators in completing tasks more effectively.  \\
   \addlinespace
\textit{H5}: Social factors $\rightarrow$ Intention to Use & Not supported. The relationship between Social factors and Intention to Use is not significant. & Although Generative AI tools are deemed important enablers of developers' self-efficacy, this perception does not translate into adoption. \\
   \addlinespace
\textit{H6}: Personal and environmental factors $\rightarrow$ Intention to Use & Not supported. This relationship is not significant. & Personal innovativeness and organizational support do not significantly influence developers' adoption of Generative AI tools. \\  
   \addlinespace
\textit{H7}: Personal and environmental factors $\rightarrow$ Perceptions about the technology & Supported. We found a significant relationship between Personal and environmental factors and Perceptions about the technology with a path coefficient of 0.31 and a small to medium effect size (0.11). & As expected, a developer's propensity to experiment with Generative AI tools, along with perceived organizational support, positively impact the perception of the technology.\\
\bottomrule
\end{tabularx}
\end{table}

Our qualitative study was foundational in the development of our framework, which stresses the impact of perceptions about the technology and compatibility factors in shaping the intention to use LLMs. However, the subsequent quantitative study upended the expectation that these perceptions directly influenced intention to use, thus contradicting the traditional Technology Acceptance Model \cite{davis1989TAM}. This suggests that, when adopting LLMs, the compatibility with existing work processes significantly impacts perceptions about the technology.

The central role of compatibility factors aligns with previous findings \cite{moore1999crossing} and reaffirms the essential need for technological alignment with current practices. Our study revealed that software engineers are more inclined to adopt LLMs when the technology fits seamlessly into their existing workflows, a finding in line with prior research \cite{dourish2003appropriation}.

In contrast to expectations, the quantitative investigation indicated that social factors did not significantly contribute to the intention to use LLMs. This deviation from previous studies \cite{venkatesh2003user} underlines the nuanced influences of social aspects and self-efficacy on the adoption process.

Personal and environmental factors, while influential in shaping perceptions about the technology, didn't directly impact the intention to use. This observation supports the assertion by \cite{agarwal1999individual} that personal innovativeness might mold perceptions about a technology but does not necessarily guarantee adoption.

The insights derived from our exploration of LLM adoption using the HACAF theory both diverge from and extend beyond established theoretical frameworks like TAM, DOI, and SCT, shedding light on the complex dynamics at play. These findings carry both academic and practical implications, underlining the necessity for holistic strategies that consider individual perceptions and compatibility factors for promoting LLM adoption.


Our research places a clear emphasis on \textbf{Compatibility Factors as primary drivers in the adoption process of Generative AI technologies}. Compatibility factors, in essence, measure how well a new technology fits into an individual or organization's existing framework—both in terms of practical workflow and broader values. They are essential in determining the successful adoption of technologies, such as Generative AI.
The importance of seamless integration within existing workflows is the core driver towards AI adoption. Workflows refer to the defined sequence of tasks that a software engineering team performs regularly, such as, coding, debugging, testing, and deployment. Notably, also hardware plays a crucial role in these processes~\cite{gao2023empirical}. If an LLM can integrate smoothly into these steps by e.g., automating certain coding tasks or improving debugging efficiency — it is seen as highly compatible. Conversely, if the integration of LLM disrupts these established procedures or necessitates significant changes to existing operations, it might be deemed less compatible, and its adoption may face resistance.
This idea extends beyond workflows to include the broader technological environment. If the LLM requires different operating systems or specific hardware that is not in use, or if it relies on knowledge or skills that the team does not possess, it can make the tool less compatible. Therefore, the technology compatibility and the alignment with existing skills are vital considerations.
In other words, even if a tool holds significant potential utility, if an individual or an organization fails to understand how it fits within their existing framework, i.e., workflow, technical environment, or value system — its adoption becomes less likely. This insight brings into focus the critical role of compatibility in designing and promoting new technology. For successful integration within the software engineering industry, Generative AI tools should be designed with a deep understanding of the existing systems, practices, and values in mind. 

Another noteworthy aspect of our findings is the non-significant relationship between Social Factors and Intention to Use, despite the positive influence of perceptions about the technology on Social Factors. This unexpected finding could be contextualized by considering the nascent stage of the Generative AI transformation within the industry. \textbf{Despite the recognition of the potential benefits and enhancements to self-efficacy offered by LLMs, this does not necessarily catalyze immediate widespread adoption, suggesting that mere perceptions about a technology's potential advantages may not be sufficient to prompt its adoption}.

Interestingly, we found that Personal and Environmental Factors did not directly lead to the Intention to Use, stressing the early stage of LLM adoption. In this initial phase, personal innovativeness and organizational support appear to exert limited influence on adoption, placing the emphasis squarely on Compatibility Factors. This suggests that \textbf{as AI tools are more seamlessly integrated within existing workflows, their likelihood of adoption increases}.

It is critical to acknowledge that \textbf{we are currently in the nascent stages of the Generative AI transformation}. As the utilization of these tools gains wider traction over time, we anticipate that the relationships within HACAF will be further corroborated. The current non-validation of all relationships within our framework model does not detract from its utility but offers initial insights into the factors shaping the adoption of Generative AI tools at this stage.

While our study primarily focused on the adoption dynamics of Generative AI tools in software engineering, it is worth noting that a minority of our informants did touch upon the \textbf{ethical implications and potential risks associated with their use}. 
Despite the limited number of informants addressing these concerns, the issues they raised are significant. Large language models, while powerful, can produce harmful outputs or inadvertently perpetuate biases present in the data they were trained on. Such outputs can have unintended consequences, especially when integrated into software products that reach a wide audience.
Recent research, including that by Tsamados et al.~\cite{tsamados2023cybersecurity}, has emphasized the security vulnerabilities inherent to artificial intelligence, especially with LLMs. Their findings suggest a need for a comprehensive understanding of these models' capabilities and vulnerabilities to ensure a balanced cost-benefit analysis. In the context of our study, it is crucial to recognize that while the compatibility of AI tools within existing development workflows is a significant driver for adoption, this should not overshadow the potential risks.
Model providers, in their journey to promote adoption, should be proactive in communicating potential vulnerabilities and offering guidance on best practices to mitigate risks. Given the rapid evolution of LLM technologies, periodic reviews and audits can ensure that models remain transparent and adhere to ethical standards. Platforms integrating LLMs should bolster their security measures, ensuring that they are equipped to handle the unique challenges posed by these models. Furthermore, when integrating or using LLMs, preference should be given to models from established and reputable sources to ensure reliability and security.
Open discussions between stakeholders, including developers, users, and regulatory bodies, can foster a more responsible and ethical adoption of Generative AI tools in software engineering. By addressing these concerns, we can ensure a balanced and responsible approach to the integration of Generative AI tools, aligning with the insights and framework provided by our study.

This study's significance lies not only in its immediate findings but also in establishing a foundation for understanding software developers' perceptions of AI during this early phase. This insight is indispensable for guiding the design and implementation of these tools to align with user needs and expectations.

Lastly, while this study offers an important snapshot of the current state of Generative AI adoption in software development, a comprehensive assessment of the HACAF model necessitates long-term and longitudinal studies. Such investigations would facilitate a more profound understanding of the evolution and interplay of the factors affecting adoption as the technology matures and gains wider adoption. Longitudinal studies will provide nuanced insights into changing adoption patterns, further refining the HACAF model over time, and contributing to a more sophisticated understanding of technology adoption phenomena.

\subsection{Implications}

The findings of this study have far-reaching implications for practitioners and researchers in software engineering and artificial intelligence, offering new insights from the perspective of the Human-AI Collaboration and Adaptation Framework (HACAF). The integration of AI, specifically Generative AI tools like Large Language Models, into software development processes has shown promise, leading to potential advancements in various aspects of the field.

A key implication of our study is \textbf{the need for organizations to consider investing in AI-driven tools that fit well within existing development workflows}. To translate this into actionable steps, we recommend:

\begin{itemize}
    \item \textit{Tool Evaluation:} Organizations should initiate a thorough evaluation of available AI-driven tools, focusing on their compatibility with current processes and the specific needs of their development teams.
    \item \textit{Pilot Testing:} Before a full-scale implementation, conduct pilot tests with a subset of projects to gauge the tool's impact on development time, software quality, and user satisfaction.
    \item \textit{Training Programs:} Introduce training sessions for developers to familiarize them with the nuances of the selected AI tools, ensuring they can leverage the tool's capabilities to the fullest.
    \item \textit{Feedback Loop:} Establish a feedback mechanism where developers can report on the tool's performance, any challenges faced, and areas of improvement. This feedback can guide iterative refinements and ensure the tool remains aligned with evolving development practices.
    \item \textit{Cost-Benefit Analysis:} Regularly assess the return on investment from the AI tool adoption, considering factors like improved efficiency, reduced errors, and user satisfaction.
\end{itemize}

Given our findings on the importance of Compatibility Factors in driving AI adoption, there is an \textbf{urgent need for continuous education and training in AI} as it becomes more prevalent in software engineering. Concretely, we suggest:

\begin{itemize}
    \item \textit{Tailored Training Modules:} Organizations should develop and offer tailored training modules that focus on the integration of AI tools within existing software development workflows. These modules should address both the technical and collaborative aspects of using AI in team settings.
    \item \textit{Regular Workshops:} Host regular workshops and seminars featuring experts in the field of AI. This will provide employees with insights into the latest advancements and best practices in AI-driven software development.
    \item \textit{Collaboration with AI Tool Providers:} Forge partnerships with AI tool providers to facilitate hands-on training sessions, ensuring that the workforce is adept at leveraging the full potential of these tools.
    \item \textit{Strengthening Industry-Academic Synergy:} To bridge the gap between theoretical knowledge and practical application, academic institutions should foster deeper collaborations with industry leaders. By co-designing software engineering curricula that emphasize AI-driven development, we can ensure that graduates are not only well-versed in current technologies but also attuned to the real-world challenges and opportunities of the industry.
    \item \textit{Feedback-Driven Iterations:} Establish mechanisms to gather feedback from employees undergoing training. This feedback can be instrumental in refining training content and methodologies, ensuring they remain relevant and effective.
\end{itemize}

Our study also underscores \textbf{the significance of the human-in-the-loop approach when incorporating AI in software engineering}. To use the full potential of AI tools while ensuring they augment, rather than replace, human capabilities, the following actionable steps are recommended:

\begin{itemize}
    \item \textit{User-Centric AI Design:} AI tool developers should prioritize a user-centric design philosophy, ensuring that the tools are intuitive and seamlessly integrate into the developers' workflow.
    \item \textit{Transparency Mechanisms:} Implement mechanisms within AI systems that elucidate their decision-making processes. This will empower developers to understand and trust the AI's suggestions and decisions.
    \item \textit{Calibration Features:} Equip AI tools with features that allow developers to calibrate or fine-tune them based on specific project requirements and their professional judgment.
    \item \textit{Feedback Loops:} Establish iterative feedback loops where developers can provide feedback on AI tool outputs, which can then be used to refine and improve the tool's performance over time.
    \item \textit{Collaborative Workspaces:} Design collaborative workspaces where both AI tools and developers can contribute in real-time, fostering a symbiotic relationship that leverages the strengths of both.
    \item \textit{Continuous Training:} As AI tools evolve, provide developers with continuous training opportunities to ensure they can effectively collaborate with the latest versions of these tools.
\end{itemize}

Importantly, the successful deployment of AI in software engineering relies heavily on aligning AI techniques with the unique needs and contexts of each organization. \textbf{A careful assessment of requirements, resources, and constraints should precede the adoption of any AI-driven solution}. To ensure that AI tools and techniques are not just adopted, but also effectively integrated, the following steps are crucial:

\begin{itemize}
    \item \textit{Requirement Analysis:} Before diving into AI adoption, organizations should conduct a thorough analysis of their specific needs. This involves understanding the challenges they face and identifying how AI can address them.
    \item \textit{Resource Evaluation:} Assess the available resources, both in terms of hardware and human expertise. This will help in selecting AI solutions that the organization has the capacity to deploy and maintain.
    \item \textit{Constraint Identification:} Recognize any constraints, be they budgetary, temporal, or technical, that might impact the deployment and operation of AI solutions.
    \item \textit{Feedback Mechanism:} Establish a mechanism for developers and other stakeholders to provide feedback on the AI tools, ensuring that they are refined and optimized over time.
    \item \textit{Continuous Review:} Periodically review the AI adoption strategy to ensure it remains aligned with the evolving needs and contexts of the organization.
\end{itemize}

Our research, while primarily centered on the adoption dynamics of Generative AI tools in software engineering, did highlight the \textbf{ethical implications and potential risks associated with their use}. Large language models, despite their capabilities, can sometimes produce outputs that might be considered harmful or inadvertently mirror biases from their training data. Such unintended outputs can have profound implications, especially when integrated into software products for a vast audience.

\begin{itemize}
    \item \textit{Security Vulnerabilities:} Recent studies, such as the one by Tsamados et al.~\cite{tsamados2023cybersecurity}, have shed light on the security vulnerabilities inherent to artificial intelligence, particularly with LLMs. Their findings underscore the need for a thorough understanding of these models' capabilities and vulnerabilities.
        \item \textit{Model Provider Responsibility:} Model providers, in their quest to foster adoption, should be proactive in detailing potential vulnerabilities and offering guidelines on best practices to counteract these risks. As LLM technologies evolve rapidly, periodic reviews and audits are paramount to ensure transparency and adherence to ethical standards.
        \item \textit{Platform Security Measures:} Platforms integrating LLMs should amplify their security protocols, ensuring they are prepared to tackle the unique challenges presented by these models. When adopting or utilizing LLMs, it's prudent to choose models from well-established and trustworthy sources, guaranteeing both dependability and security.
        \item \textit{Stakeholder Dialogues:} Open conversations among stakeholders, encompassing developers, users, and regulatory entities, can champion a more responsible and ethical adoption of Generative AI tools in software engineering.
\end{itemize}

In summary, the integration of AI techniques, especially Generative AI tools like LLMs, in software engineering holds immense potential for enhancing the development process and improving overall software quality. By acknowledging and addressing the challenges associated with AI adoption—taking into consideration our findings on the importance of Compatibility Factors—organizations can effectively leverage AI-driven tools and methodologies to realize enhanced outcomes in their software development pursuits.

\subsection{Limitations}

Our threats to validity are deliberated in relation to the application of both qualitative and quantitative validity frameworks, as advised by earlier research~\cite{Russo2018ISQ,russo2021agile}. Consequently, we commence our discourse on the credibility, transferability, and confirmability for qualitative examination~\cite{Guba}.

\textbf{Credibility}. The principal variables of our investigation were informed by the three core theoretical models that assess individual, technological, and social-level influences. These models are the Technology Acceptance Model, the Diffusion of Innovation Theory, and the Social Cognitive Theory. By integrating these thoroughly validated theories into our study, we could deeply understand the factors influencing language model adoption, and further probe how these variables are implemented within the software engineering domain. Moreover, to ensure the credibility of our data, informants underwent a rigorous multistage selection process, verifying their roles as software engineers actively working with Generative AI tools. This meticulous selection process fortified the integrity of our study and the resulting insights.

\textbf{Transferability and Confirmability}. Our qualitative data analysis was performed by a single researcher, applying the Gioia Methodology to the collected data. The utilization of a single researcher for data analysis could be perceived as a limitation due to potential biases and subjectivity. However, employing the Gioia Methodology significantly mitigates these biases, contributing to the trustworthiness of our findings. The Gioia methodology provides a structured approach that emphasizes transparency, iterative categorization, and systematic data processing, which lends itself to limiting subjective bias~\cite{gioia2013seeking}.
Moreover, our research merged qualitative data with a sample study to bolster the transferability of our findings. Transferability, as defined in qualitative research, refers to the extent to which the results can be applied in other contexts or with other respondents~\cite{Creswell}. By incorporating a sample study, we provided a broader base from which parallels could be drawn, thereby enhancing the applicability of our research to varied contexts.
Despite surveying a broad group of participants, limitations were present as we were unable to conduct follow-up inquiries. This may potentially affect the depth and comprehensiveness of our data. Yet, our utilization of the Gioia Methodology and the combination of qualitative and sample study data have fortified the structure and interpretative robustness of our data.
\\

Furthermore, we disucss statistical conclusion, internal, construct, and external validity to assess the quantitative investigation~\cite{Wohlin}.

\textbf{Internal}. Our research model was validated using a cluster-randomized probability sampling strategy~\cite{gravetter2018research}. Because of feasibility issues, we selected a cluster of the global population (i.e., the Prolific community) instead of the entire population. Although less accurate than random sampling, cluster sampling is more cost-efficient. In response to Baltes and Ralph's call, which noted that less than 10\% of software engineering studies in top venues utilize probability sampling~\cite{baltes2020sampling}, we designed our study accordingly. Our data quality was boosted by a multi-stage process in which only 184 carefully selected professionals were chosen out of the 831 potential candidates identified (approximately 22\% of the initial candidates). Nevertheless, our sample is not representative of the software engineering population.

\textbf{External}. The generalization of our findings was a significant concern during the PLS-SEM analysis as sample studies are best suited for theory generalization~\cite{stol2018abc}. We gathered 184 responses, an ample size considering the \textit{a priori} power study we conducted prior to data collection.

\textbf{Construct}. Constructs were gauged via a single-informant approach, embodying a software engineer's viewpoint. Additionally, we employed self-reported measures, asking participants to express their agreement level on literature-derived indicators. However, these questions might not have been accurately answered. To counteract these limitations, we introduced three random attention checks, which eleven candidates failed. Furthermore, we adjusted our measurement instrument based on pre-existing ones. Lastly, we randomized the questionnaire and tested it for clarity and consistency to manage potential accuracy biases.

\textbf{Statistical conclusion}. We processed the survey results using Partial Least Squares -- Structural Equation Modelling with the renowned statistical software SmartPLS (4.0.9.5), which has been utilized in over 1,000 peer-reviewed articles~\cite{ringle2015smartpls}. All statistical methods and tests employed for the PLS-SEM analysis are in line with the most recent guidelines in our field~\cite{russo2021pls}.

\section{Conclusion}
\label{sec:conclusion}

This study presents a mixed-methods investigation about the adoption of Generative AI tools within the field of software engineering. By developing the Human-AI Collaboration and Adaptation Framework (HACAF), our research provides a nuanced understanding of the complexities involved in the adoption of such technologies, particularly during these nascent stages of the Generative AI transformation.

Our findings shed light on the pivotal role of Compatibility Factors, emphasizing the need for AI tools to fit within existing development workflows to enhance their adoption. This points towards the understanding that adoption is not driven solely by the perceived benefits of the technology, but by its seamless integration into the user's pre-established work processes.

Beyond its theoretical contributions, this study has significant practical implications. It offers early insights into software developers' perceptions of AI, providing valuable pointers for the design and refinement of user-focused AI tools. These insights can help foster a more widespread adoption of AI tools by addressing developer concerns and optimizing tool compatibility with existing workflows.

As we look forward, future research in the nascent field of AI and software engineering can build upon the foundation laid by this study. 
While this study delivers an important snapshot of the current state of Generative AI adoption in software development, it is crucial to recognize that a comprehensive assessment of the HACAF model necessitates long-term and longitudinal studies. Such investigations would permit a deeper understanding of the evolution and interplay of factors influencing adoption as the technology matures and gains broader acceptance. These longitudinal studies would offer nuanced insights into the changing adoption patterns, thus allowing the continuous refinement of the HACAF model and contributing to a more sophisticated understanding of technology adoption phenomena.


\section*{Supplementary Materials} 
The PLS-SEM computational tables, raw data, the survey instruments, and the overfitting analysis are openly available under a CC BY 4.0 license on Zenodo, DOI: \url{https://doi.org/10.5281/zenodo.8124332}.

\section*{Acknowledgment}
This work was supported by the The Danish Industry Foundation with the Sb3D project --- Security by Design in Digital Denmark.

ChatGPT-4 has been used to ensure linguistic accuracy and enhancing the readability of this article.

\bibliographystyle{ACM-Reference-Format}
\bibliography{bib}

\appendix

\section{Appendix A (Survey Instrument)}
\label{app:Appendix}

\begin{table}[ht!]
\centering
\sisetup{
group-digits=true,
group-minimum-digits=4,
table-format=0.3,
mode=text,
detect-weight=true,
detect-family=true
}
\footnotesize
\robustify{\textbf}
\caption{Items description. Those prefixed with (*) were dropped because of their insufficient loading onto their latent variable}
\label{tab:Items}
\begin{tabular}{p{2cm} p{1cm} p{8cm} p{1cm}}
\toprule
Construct &
Item ID &
Questions &
Reference  \\
  \midrule
 \\
Perceptions about the Technology & PT\_1
&
Using LLMs in my job enables me to accomplish tasks more quickly. & \cite{davis1989TAM,venkatesh2003user}
\\
& PT\_2
&
Using LLMs would improve my job performance. & \cite{davis1989TAM,venkatesh2003user}
\\
& PT\_3 & Using LLMs in my job would increase my productivity. & \cite{davis1989TAM,venkatesh2003user} \\
& PT\_4
&
Using LLMs would enhance my effectiveness on the job. & \cite{davis1989TAM,venkatesh2003user}
\\
& PT\_5
&
  (*) Using LLMs would make it easier to do my job. & \cite{davis1989TAM,venkatesh2003user}
\\
& PT\_6
&
I would find LLMs useful in my job. & \cite{davis1989TAM,venkatesh2003user}
\\
& PT\_7
&
I would find LLMs easy to use. & \cite{davis1989TAM,venkatesh2003user}
\\
& PT\_8
&
  (*) It would be easy for me to become skillful at using LLMs. & \cite{davis1989TAM,venkatesh2003user}
\\
& PT\_9
&
  (*) I would find LLMs flexible to interact with. & \cite{davis1989TAM,venkatesh2003user}
\\
& PT\_10
&
  (*) Learning to operate LLMs would be easy for me. & \cite{moore1999crossing,venkatesh2003user}
\\
& PT\_11
&
Using LLMs would enable me to accomplish tasks more quickly. & \cite{davis1989TAM,venkatesh2003user}
\\
& PT\_12
&
Using LLMs would improve the quality of work I do. & \cite{moore1999crossing,venkatesh2003user}
\\
& PT\_13
&
Using LLMs would make it easier to do my job. & \cite{moore1999crossing,venkatesh2003user}
\\
& PT\_14
&
Using LLMs would enhance my effectiveness on the job. & \cite{moore1999crossing,venkatesh2003user}
\\
& PT\_15
&
  (*) Using a LLM gives me greater control over my work. & \cite{ajzen1991theory,venkatesh2003user}
\\
\addlinespace
Compatibility Factors & CF\_1
&
  (*) Using LLMs is compatible with all aspects of my work. & \cite{moore1999crossing,venkatesh2003user}
\\
& CF\_2
&
I think that using a LLM fits well with the way I like to work. & \cite{moore1999crossing,venkatesh2003user}
\\
& CF\_3
&
Using a LLM fits into my work style. & \cite{moore1999crossing,venkatesh2003user}
\\
& CF\_4
&
  (*) I have control over using LLMs. & \cite{ajzen1991theory,venkatesh2003user}
\\
& CF\_5
&
I have the knowledge necessary to use LLMs. & \cite{ajzen1991theory,venkatesh2003user}
\\
& CF\_6
&
Given the resources, opportunities, and knowledge it takes to use the LLMs, it would be easy for me to use LLMs. & \cite{ajzen1991theory,venkatesh2003user}
\\
\addlinespace
Social Factors & SF\_1
&
People who influence my behaviour think that I should use LLMs. & \cite{ajzen1991theory,venkatesh2003user}
\\
& SF\_2
&
People who are important to me think that I should use LLMs. & \cite{ajzen1991theory,venkatesh2003user}
\\
& SF\_3
&
  (*) I use LLMs because of the proportion of coworkers who use the system. & \cite{thompson1991personal,venkatesh2003user}
\\
& SF\_4
&
  (*) People in my organisation who use LLMs have more prestige than those who do not. & \cite{moore1999crossing,venkatesh2003user}
\\
\addlinespace
Personal and Environmental Factors & PEF\_1
&
  (*) If I heard about a new technology like LLMs, I would look for ways to experiment with it. & \cite{venkatesh2003user}
\\
& PEF\_2
&
  (*) Among my peers, I am usually the first to try out new technologies like LLMs. & \cite{agarwal1998conceptual}
\\
& PEF\_3
&
  (*) I like to experiment with new technologies.  & \cite{agarwal1998conceptual}
\\
& PEF\_4
&
My organization provides the necessary resources for using LLMs effectively. & \cite{kulkarni2006knowledge,venkatesh2003user}
\\
& PEF\_5
&
My organization offers sufficient training sessions to enhance our skills in using LLMs. & \cite{kulkarni2006knowledge,venkatesh2003user}
\\
& PEF\_6
&
My organization encourages the use of LLMs in our daily work. & \cite{lewis2003sources,venkatesh2003user}
\\
& PEF\_7
&
I feel that there is top management support in my organization for using LLMs. 
& \cite{lewis2003sources,venkatesh2003user}
   \\
   \addlinespace
Intention to Use & IU\_1
   &
  I intend to use LLMs more extensively in the next months. & \cite{davis1989TAM,venkatesh2003user}
   \\
 & IU\_2
   &
  I predict I would use LLMs more extensively in the next months. & \cite{davis1989TAM,venkatesh2003user}
   \\
 & IU\_3
   &
  I plan to use LLM more extensively in the next months. &   \cite{davis1989TAM,venkatesh2003user} \\
   \addlinespace
   \midrule
\end{tabular}
\end{table}

\newpage

\section{Appendix B (Overfitting Analysis)}
\label{sec:Overfitting}

The notably high effect size observed between `Perception about the Technology' and `Compatibility Factors' might raise concerns of potential overfitting, a scenario where the model inadvertently learns noise in the training data, thereby compromising its ability to accurately predict unseen data \cite{hawkins2004problem}. Given the potential implications of overfitting on the reliability of our model, it is crucial to thoroughly investigate this issue.
The details of this analysis, along with the associated code and data, can be found in the online supplementary materials hosted on Zenodo\footnote{Link to the replication package: https://doi.org/10.5281/zenodo.8124332.}. 

First, we analyzed the residuals of our model, which can provide insights into the appropriateness of the model fit. The residuals represent the difference between the observed and predicted values for the dependent variable. In a well-specified model, we would expect the residuals to be randomly scattered around zero, with no apparent pattern. This would suggest that the model's errors are random, and that the model is correctly specified.

\begin{figure}[h]
\centering
\includegraphics[width=0.7\textwidth]{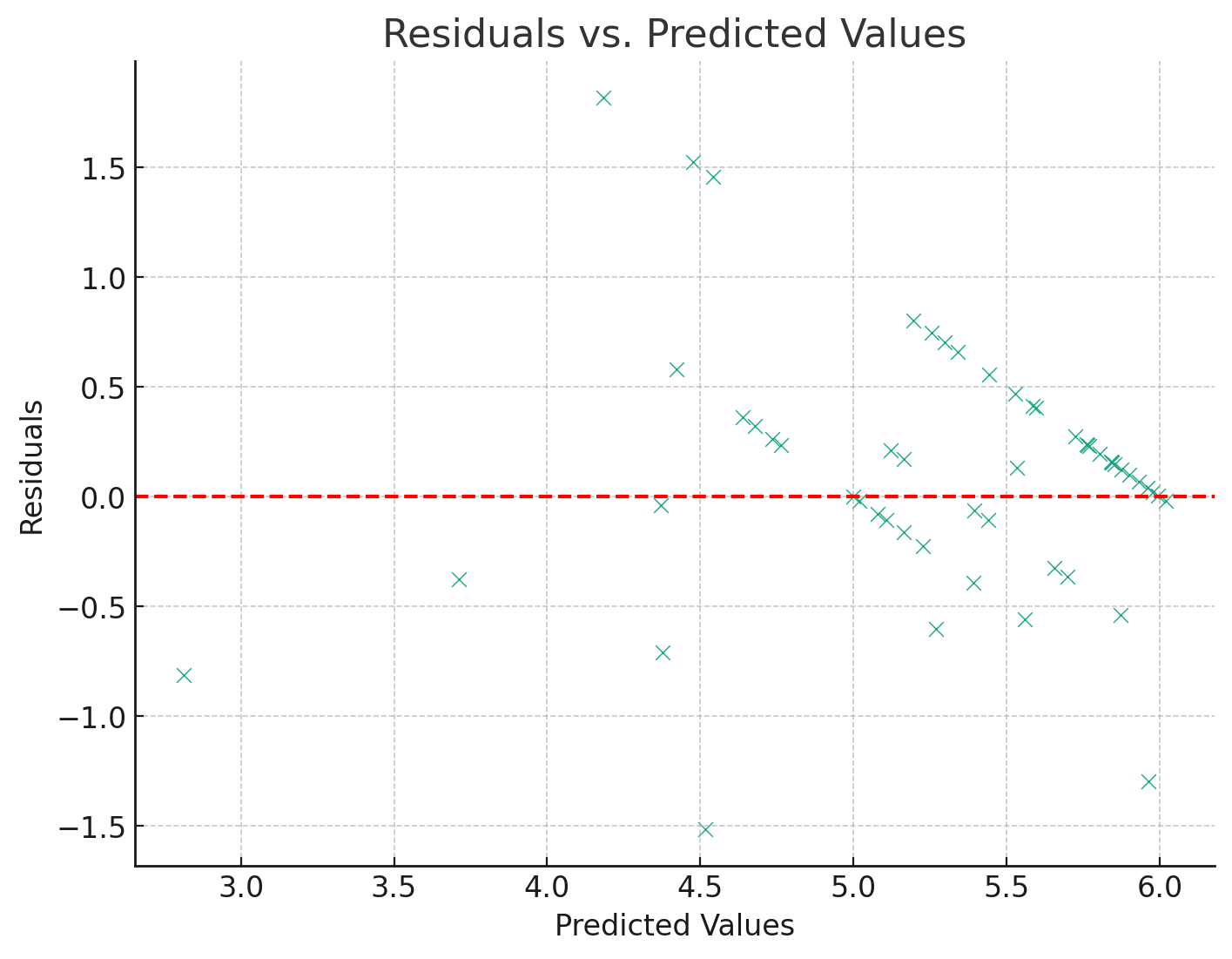}
\caption{Residuals vs. predicted values for the `Intention to Use' construct. The residuals are mostly randomly scattered around zero, suggesting that the model's errors are random and that the model is correctly specified.}
\label{fig:residuals}
\end{figure}

We plotted the residuals against the predicted values for the 'Intention to Use' construct and found that they were indeed mostly randomly scattered, suggesting that our model's errors are random. Figure \ref{fig:residuals} shows this residuals plot.

However, to evaluate the risk of overfitting more thoroughly, we employed two key methodologies: a train-test split and cross-validation \cite{kohavi1995study}. In the train-test split, we partitioned our data into a training set (80\% of the data) and a testing set (20\% of the data). Our model was trained on the training data and then evaluated on the unseen testing data. 

The model's performance was assessed using the mean squared error (MSE), a metric that calculates the average squared difference between the observed and predicted values. A lower MSE indicates a better fit to the data. The MSE for the test set was approximately 0.523. 

To further examine overfitting, we implemented k-fold cross-validation with k set to 5, a common choice for this parameter \cite{kohavi1995study}. In this approach, the data was divided into 5 subsets, and the model was trained and tested 5 times, each time on a different subset of the data. The performance of the model was again assessed using the MSE. The mean MSE from the cross-validation was approximately 0.676, reasonably close to the test MSE, suggesting that our model is not overfitting the data.

In conclusion, both the train-test split and cross-validation results suggest that our model is generalizing effectively to unseen data and is not overfitting.

\end{document}